\documentclass[12pt,a4paper]{article}
\usepackage{amsmath}
\usepackage{amssymb}
\usepackage{braket}
\usepackage[pdftex]{graphicx}
\usepackage[pdftex]{hyperref}
\usepackage[titletoc,title]{appendix}
\usepackage{url}

\numberwithin{equation}{section}
\allowdisplaybreaks[1]

\title{Exact large deviation function of spin current for the one-dimensional XX spin chain with domain wall initial condition}
\author{H. Moriya\footnote{hmoriya@stat.phys.titech.ac.jp} , 
            R. Nagao\footnote{r-nagao@stat.phys.titech.ac.jp}, 
            \vspace{2mm}
            T. Sasamoto\footnote{sasamoto@phys.titech.ac.jp} \\
            Department of Physics, Tokyo Institute of Technology, \\
             2-12-1~Ookayama, Meguro-ku, Tokyo 152-8550, Japan}
\date{}

\begin{document}

\maketitle

\begin{abstract}
We investigate the fluctuations of the spin current for the one-dimensional XX spin chain starting from the domain wall initial condition.
The generating function of the current is shown to be written as a determinant with the Bessel kernel.  
An exact analytical expression for the large deviation function is obtained by applying the Coulomb gas method. 
Our results are also compared with DMRG calculations.
\end{abstract}

\section{Introduction}

Recently, quantum dynamics of many-body systems attract interests of many researchers, and there has been impressive progress in both experiments and theories \cite{bloch2008many,polkovnikov2011colloquium}.
It is generally hard to study dynamical properties of quantum many-body systems analytically.
But in the case of one dimension, some exact results have been obtained.
They have not only provided us useful theoretical insights about quantum dynamics and thermalization but are also in many cases experimentally relevant.

For example, the famous experiment about a quantum version of Newton's cradle with one-dimensional Bose gases \cite{kinoshita2006quantum} proposed an example of systems which do not thermalize due to their integrability.
This experiment inspired the introduction of the generalized Gibbs ensemble (GGE) \cite{rigol2007relaxation} as a state quantum integrable systems tend to relax to.

More recently, a hydrodynamic formulation called the Generalized Hydrodynamics (GHD) was proposed in \cite{castro2016emergent,bertini2016transport} and has been successfully applied to various quantum integrable systems such as XXZ Heisenberg chain, Lieb-Liniger model, for calculating their density, current profile and so on.

One of the most standard setups to study systems out of equilibrium is the so-called partitioning protocol in an isolated quantum system, in which initially two regions of a system are prepared in macroscopically different states, at $t=0$ they are connected abruptly, and then one studies the time evolution of the whole system.

In the context of spin chains, the simplest partitioning protocol is to take the domain wall initial state, in which spins in two regions are prepared in the opposite directions, and to study the spin dynamics after that.
This inhomogeneous magnetization profile before the quantum quench takes place helped to reveal several non-equilibrium behaviors in different models (see for instance \cite{bertini2016transport,eisler2016universal,lancaster2016nonequilibrium}).
Domain wall initial state having a step-like spin configuration is pretty easy to make in real materials because we only have to apply the external strong magnetic field into each system.

Many results have been obtained for this setting.
For example, for the case of the XX spin chain, the magnetization density profile has already been found in \cite{antal1999transport}.
It is also reproduced by GHD in \cite{bertini2016transport}.
Time-dependent DMRG were also applied to spin chains' dynamics initiated from this condition (for instance \cite{gobert2005real,ljubotina2017spin,misguich2017dynamics}) including different anisotropy parameter $\Delta$ of XXZ spin chain.
But most results so far have been for average behaviors of the systems.
Fluctuation properties have much less been addressed, though very recently there was important progress in the context of GHD \cite{myers2018transport}.

In this paper, we focus on the XX chain and integrated spin current $N(t)$ which is given by the total number of up spins which moved from the left to the right subsystem for enough large time period $t$ under the domain wall initial condition at absolute zero temperature.

The large time behavior of the mean and the variance of this quantity have been already known in the previous works \cite{antal1999transport,antal2008logarithmic} :
\begin{equation}
\braket{N(t)}\simeq\frac{t}{\pi},\quad\braket{N(t)^2}_c=\braket{\left[N(t)-\braket{N(t)}\right]^2}\simeq\frac{1}{2\pi^2}(\log{t}+C),
\label{eq:AveandVar}
\end{equation}
where the constant $C$ is given as a sum of a few constants and definite integrals and numerically $C=2.963 510 026\dots$.
The notation $\simeq$ means that both sides are asymptotically equal.
\noindent
Our aim in this paper is to study the full distribution of $N(t)$ for large $t$, in particular to give an explicit formula for the large deviation function of $N(t)$.

The XX model is well-known to be equivalent to a free fermionic system, for which the seminal Levitov-Lesovik formula \cite{levitov1993charge} is available.
The formula, in particular, the long-time approximated form of it (see eq. (9) in \cite{levitov1993charge}) was established in 1993 by using the scattering matrix and has a wide application, but for our setting with zero temperature and perfect transmission starting from the domain wall state, a naive application of the Levitov-Lesovik formula would lead to the Dirac delta shape distribution and does not create the variance in eq. (\ref{eq:AveandVar}) which increases logarithmically in time as pointed out in \cite{schonhammer2007full,schonhammer2009full}.

In this paper, we do not rely on the long-time approximated Levitov-Lesovik formula and study the statistics of $N(t)$, in particular the large deviation function (a.k.a. the rate function), by exact calculations.
It contains far more information than the average and the variance in eq. (\ref{eq:AveandVar}).
As is well known, the moment generating function and its large deviation function are mutually convertible through the Legendre transform \cite{dembo1998large}.
The large deviation in non-equilibrium steady state (NESS) has been investigated for open systems with boundary driven condition \cite{vznidarivc2014large,vznidarivc2014exact,carollo2018current}, however, we directly treat the isolated system from the quench without assuming NESS.

Our arguments are based on the determinant formula for the generating function of $N(t)$ found in \cite{eisler2013full}.
In \cite{eisler2013full}, the authors mainly studied the quantum propagating front which is the edge of the melting up spins moving to the other subsystem.
They found that the statistics at the quantum front is described by the Airy kernel \cite{forrester1993spectrum} asymptotically and thus showed that the rightmost up spin's existence probability is given by the Tracy-Widom distribution \cite{tracy1994levelA}.
In this paper, we will rewrite the generating function in terms of the Bessel kernel, and apply the Coulomb gas technique to study the large deviation properties.
We also found that time-dependent DMRG method can be extended to calculate the rate function through moment generating function by applying matrix product operators having the dependence on $\lambda$ successively.

The rest of the paper is organized as follows.
In section 2, we introduce the model and some notations and also present our main result.
In section 3, we explain the mapping from our dynamical problem into a static problem in terms of random matrix theory.
Section 4 is to be devoted to a review of the role of the Bessel kernel \cite{forrester1993spectrum} in random matrix theory which appeared in the previous section.
In section 5, we will explain the techniques to study the large deviation properties of the Wishart matrix and apply it to our setting to find the exact formula for the large deviation function of $N(t)$.
In section 6, by expanding the large deviation function around the average, we reproduce the variance which increases logarithmically in time.
In section 7, we will apply the time-dependent DMRG calculation to the evaluation of the large deviation function.
Section 8 is the conclusion. 

\section{The model and results}

\begin{figure}
\centering
\includegraphics[width=10cm,clip]{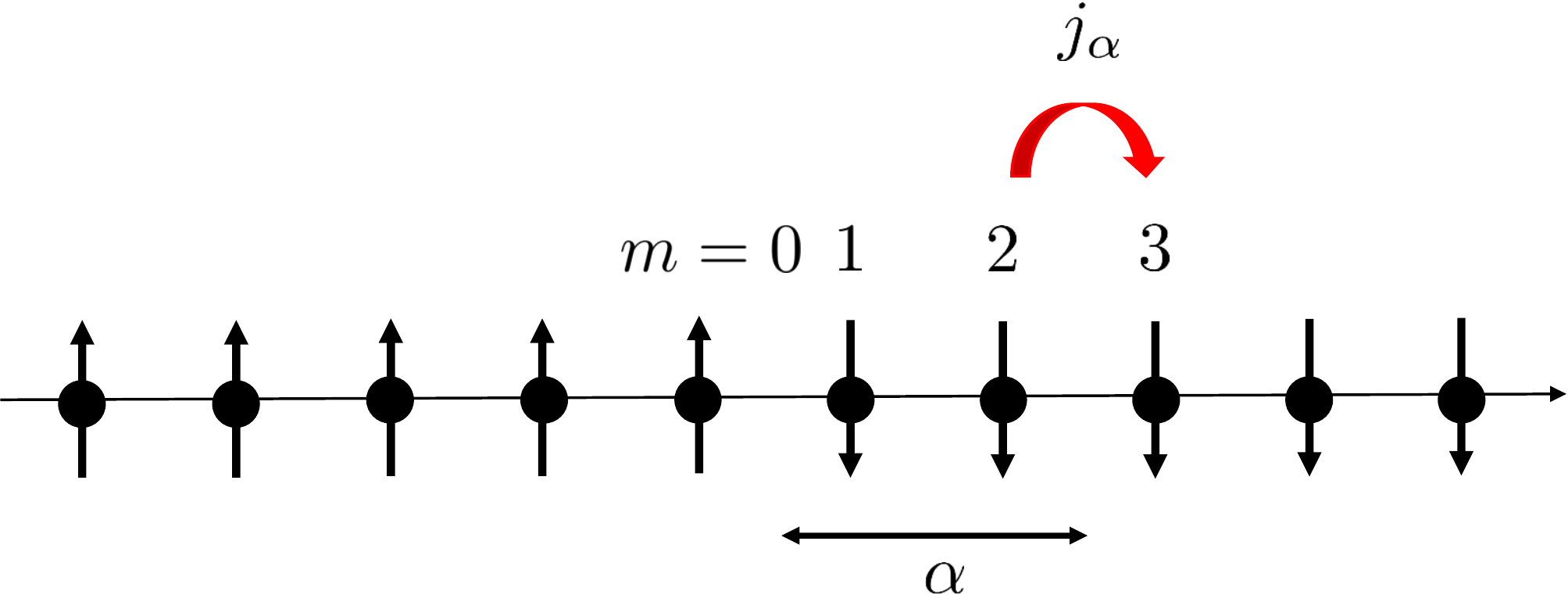}
\caption{The model. At the beginning $t=0$, the left half of the chain is fully filled by up spins and the right half of the chain is filled by down spins. In this picture $\alpha=2$, meaning that we consider the current of up spins between the sites $\alpha=2$ and $\alpha+1=3$.}
\label{fig:model}
\end{figure}

We consider the dynamics of the XX model on the one-dimensional lattice. 
A site at which a spin is located is designated by an integer. 
The Hamiltonian is given by
\begin{equation}
H_{XX}=-J\sum_{m=-\infty}^\infty (s_m^xs_{m+1}^x+s_m^ys_{m+1}^y).
\end{equation}
Here,  $J$ is the coupling constant and $s_m^i$ denotes the spin operator at site $m$ defined as half of the Pauli operator,
\begin{equation}
s_m^i=\sigma_m^i/2,\quad m\in\mathbb{Z},\quad i=x,y,z.
\end{equation}
As an initial condition at $t=0$, we employ the domain wall initial condition, in which for the left half of the lattice $m\le0$, all the spins are set to be up spins, while for the right half of the lattice $m\ge1$, all the spins are set to be down spins (see Fig.\ref{fig:model}).
We denote this state by 
\begin{equation}
\ket{\mathrm{DW}}=\ket{\cdots\uparrow\uparrow\uparrow\downarrow\downarrow\downarrow\cdots}.
\label{DW}
\end{equation}
For this model, the quantity we are interested in is the integrated spin current, which is defined as 
\begin{equation}
N(t):=\sum_{m=\alpha+1}^\infty \left[s_m^z(t)+\frac{1}{2} \right].
\label{eq:spinnum}
\end{equation}
Here, the time dependence of an operator is from the Heisenberg picture.
Without loss of generality, we can only consider the case $\alpha\ge0$ because, as long as the domain wall initial condition is considered, we can use the symmetry of the system between up spins and down spins.
Counting up spins, for example, in the region $m\ge-1$ or $\alpha=-2$ is equivalent to counting down spins in the region $m\le-2$.

To see the meaning of this observable $N(t)$, let us consider the total magnetization  operator in the area $m>\alpha$ as
\begin{equation}
M(t):=\sum_{m=\alpha+1}^\infty s_m^z(t).
\label{eq:totmag}
\end{equation}
For our domain wall initial condition, this quantity is infinite, but the difference $M(t)-M(0)=N(t)$, which can be interpreted as the difference of the total magnetization in the area $m>\alpha$ between $0$ and $t$ is well-defined.
At time $t=0$, we can, of course, confirm that $N(0)=0$ from the definition of the initial condition. 

When $\alpha=0$, we are supposed to count all the transported up spins in the right half of the chain $m\ge1$ at $t$.
Note that the dependence on $\alpha$ is abbreviated in the above notations of $N(t)$ and $M(t)$.
From the time derivative of the local magnetization $s_m^z$, 
\begin{equation}
\frac{d}{dt}s_m^z(t)=i\left[H_{XX},s_m^z\right]=j_{m-1}-j_m,
\label{eq:D_ts_z}
\end{equation}
we can also consider the instantaneous spin current \cite{antal1998isotropic},
\begin{equation}
j_m=J\left(s_{m+1}^xs_m^y-s_{m+1}^ys_m^x\right).
\label{eq:j_n}
\end{equation}
Here and in the following, we set the reduced Planck constant $\hbar$ to be unity for simplicity.

In terms of the time-dependent spin current $j_\alpha(t)$, the integrated current $N(t)$ given in eq. (\ref{eq:spinnum}) can also be written as
\begin{equation}
N(t)=\int_0^tj_\alpha(\tau)d\tau.
\label{eq:intofj}
\end{equation}
This can be immediately derived by taking the sum of the local conservation law in eq. (\ref{eq:D_ts_z}) from $m=\alpha+1$ to $\infty$ and integrating with regard to time from $0$ to $t$.
Here, we implicitly assumed the situation that the effect of the other boundary current away from the center of the chain is zero,
\begin{equation}
\lim_{m\to\pm\infty}j_m(t)=0.
\end{equation}

To consider the statistics of the total number of transmitted up spins $N(t)$, we define the moment generating function as
\begin{equation}
\chi(\lambda,t):=\braket{e^{\lambda N(t)}}.
\label{eq:mgf}
\end{equation}

The moment generating function can be expanded as
\begin{equation}
\braket{e^{\lambda N(t)}}=\sum_{n=0}^\infty e^{\lambda n}\mathrm{P}[N(t)=n].
\label{eq:mgfexp}
\end{equation}
Here $\mathrm{P}[N(t)=n]$, defined as 
\begin{equation}
\mathrm{P}[N(t)=n]=\sum_{\substack{s_m^z=\pm1/2,~m\in\mathbb{Z} \\ \\ N:=\sum_{m=\alpha+1}^\infty(s_m^z+1/2)=n}}\braket{\mathrm{DW}|e^{itH_{XX}}|\dots,s_m^z,\dots}\braket{\dots,s_m^z,\dots|e^{-itH_{XX}}|\mathrm{DW}},
\label{eq:defP}
\end{equation}
has the meaning of the probability that the total number of transported up spins $N(t)$ at $t$ is $n$.
The letter $N(t)$ is slightly abused because here it is treated like a random variable, not an operator as defined in eq. (\ref{eq:spinnum}), but there should be no confusion.
Here we remark that the probability $\mathrm{P}[N(t)=n]$ is, in fact, independent of the sign of the coupling constant $J$, because changing the sign of $J$ is equivalent to reversing the time, the Hamiltonian $H_{XX}$ is real symmetric and the domain wall initial state has only real components.
In other words, the distribution of the spin current $N(t)$ is actually the same for ferromagnetic and antiferromagnetic chains.
Therefore in the following discussions, we set $J>0$ for convenience.

The probability $\mathrm{P}[N(t)=n]$ can be recovered from the moment generating function $\chi(\lambda,t)$.
After the replacement of $\lambda$ by $i\lambda$, the Fourier coefficients of the characteristic function $\chi(i\lambda,t)$ give the probability $\mathrm{P}[N(t)=n]$ for each $n$.

In this paper, our aim is to calculate the large deviation function of the spin current exactly under the specific setup explained above.
Our main result is the following.
For the case of $\alpha=0$, the full distribution function of  the integrated spin current behaves for long time as
\begin{equation}
\mathrm{P}[N(t)=aJt]\sim e^{-t^2J^2\Psi(a)}
\label{eq:IntroResult}
\end{equation}
where the large deviation function can be described in a parametric way $\Psi(a)=F(r(a))$ as
\begin{equation}
F(r)=\frac{1}{4}(1-r)+a(r)\sqrt{r}\left[K\left(1-\frac{1}{r}\right)-E\left(1-\frac{1}{r}\right)\right]
\end{equation}
with
\begin{equation}
a(r)=
\begin{cases}
\quad\dfrac{1}{\pi}\sqrt{1-r}\left[E\left(\dfrac{r}{r-1}\right)-K\left(\dfrac{r}{r-1}\right)\right], &\mathrm{for}\quad 0\le r<1, \\[4ex]
\quad\dfrac{1}{\pi}\sqrt{r}E\left(\dfrac{1}{r}\right), &\mathrm{for}\quad 1<r.
\end{cases}
\end{equation}
and $r(a)$ is the inverse function of $a(r)$.
Here $E$ and $K$ are the complete elliptic integrals of the first and second kind, see eq. (\ref{eq:KandE}) for their definitions, and $\sim$ means that both hand sides are equal after their logarithm taken, divided by $t^2$ and performing the limit $t\rightarrow\infty$.

To solve our problem, we use a well-known mapping from the XX spin chain to a free fermion.
Let us introduce the raising and lowering operators acting on each Hilbert space $\mathbb{C}^2$ at site $m$,
\begin{equation}
s_m^\pm:=s_m^x\pm is_m^y.
\end{equation}
Performing the Jordan-Wigner transformation,
\begin{align}
c_j^\dagger &= s_j^+\exp\left(+i\pi\sum_{k}^{j-1}s_k^+s_k^-\right)=s_j^+\prod_{k}^{j-1}(1-2s_k^+s_k^-)=s_j^+\prod_{k}^{j-1}(-\sigma_k^z), \\
c_j &= s_j^-\exp\left(-i\pi\sum_{k}^{j-1}s_k^+s_k^-\right)=s_j^-\prod_{k}^{j-1}(1-2s_k^+s_k^-)=s_j^-\prod_{k}^{j-1}(-\sigma_k^z),
\end{align}
one can map a spin chain into a fermionic chain.
For the XX spin chain, one obtains the free fermionic Hamiltonian :
\begin{equation}
H_{XX}=-\frac{J}{2}\sum_{m=-\infty}^\infty\left(c_m^\dagger c_{m+1}+c_{m+1}^\dagger c_m\right)=\sum_{i,j=-\infty}^\infty c_i^\dagger H_{ij}c_j,
\label{eq:fermiHxx}
\end{equation}
where the single particle Hamiltonian is just given by
\begin{equation}
H_{ij}=-\frac{J}{2}(\delta_{i,j-1}+\delta_{i,j+1}).
\label{eq:singleH}
\end{equation}
In the following, the ferromagnetic coupling constant $J$ is set to unity.
Using the eq. (\ref{eq:fermiHxx}), the time evolution of fields are as follows \cite{antal1999transport},
\begin{equation}
c_m(t)=\sum_{n=-\infty}^\infty U_{mn}(t,0)c_n=\sum_{n=-\infty}^\infty i^{m-n}J_{m-n}(t)c_n,
\label{eq:timeevolution1}
\end{equation}
where $U$ is the time evolution operator whose matrix element is given in eq. (\ref{eq:timeevolution2}).
Here, $J_n(t)$ is the Bessel function of the first kind of order $n$ \cite{olver2010nist}, generally defined by
\begin{equation}
J_\nu(z)=\sum_{k=0}^\infty\frac{(-1)^k}{k!\Gamma(k+\nu+1)}\left(\frac{z}{2}\right)^{2k+\nu}.
\end{equation}
In the free fermion language, the state corresponding to the domain wall initial condition in eq. (\ref{DW}) is written as
\begin{equation}
\ket{\mathrm{DW}}=\prod_{m=-\infty}^0c_m^\dagger\ket{0},
\end{equation}
where $\ket{\mathrm{0}}$ means the vacuum state which does not contain any fermion.
The total magnetization flowing into the region $m>\alpha$ in eq. (\ref{eq:spinnum}) equals the number of the transmitted charges,
\begin{equation}
N(t)=\sum_{m=\alpha+1}^\infty c_m^\dagger(t)c_m(t).
\end{equation}
The spin current in eq. (\ref{eq:j_n}) also has a fermionic representation,
\begin{equation}
j_m=-\frac{1}{2i}\left(c_{m+1}^\dagger c_m-c_m^\dagger c_{m+1}\right).
\end{equation}

\section{Full counting statistics and random matrix analogy}
\label{sec:FCSandRMT}

In the studies of quantum transport, especially in the field of mesoscopic physics, obtaining the full distribution of the transferred charges is of great interest \cite{nazarov2009quantum}.
This is referred to as full counting statistics (FCS) in literature.
The moment generating function $\chi(\lambda,t)$ of the number of transferred fermions $N(t)$ in this context has been investigated in a series of studies \cite{levitov1993charge,levitov1996electron,klich2002full,schonhammer2007full,schonhammer2009full} and many others.
For the case of free fermion at zero temperature $T=0$, it is known to be written in a form of determinant,
\begin{equation}
\chi(\lambda,t)=\mathrm{det}\left[1+(e^{\lambda}-1)C(t)\right]_{\ell^2(\mathbb{Z}_{>\alpha})},
\label{eq:mgfwithC}
\end{equation}
where $C(t)$ is an infinite dimensional matrix.
When there is no fermion on the right chain, its element is of the form
\begin{equation}
C_{mn}(t)=\braket{c_m^\dagger(t)c_n(t)}=\frac{1}{2\pi}\int_{-k_F}^{k_F}\varphi_m^\ast(k;t)\varphi_n(k;t)dk.
\end{equation}
Here $\varphi_m(k;t)$ is the wave function at the position $m$ and at time $t$.
Letter $k$ denotes the wave number in the single particle system and $k_F$ is its Fermi  wave number.
In our specific case, considering the unitary time evolution in eq. (\ref{eq:timeevolution1}), the explicit form of $\varphi_m(k;t)$ is given by
\begin{equation}
\varphi_m(k;t)=\sum_{j=-\infty}^0i^{m-j}J_{m-j}(t)e^{ikj},
\end{equation}
with $-\pi<k\le\pi$ (i.e. $k_F=\pi$).
After some calculations, as shown in \cite{eisler2013full}, the moment generating function can be written as
\begin{equation}
\chi(\lambda,t)=\mathrm{det}\left[1+(e^{\lambda}-1)K_\mathrm{DBes}(t)\right]_{\ell^2(\mathbb{Z}_{>\alpha})},
\label{eq:FDDBes}
\end{equation}
where the discrete Bessel kernel \cite{borodin2000asymptotics,johansson2001discrete} is given by
\begin{equation}
\begin{split}
(K_\mathrm{DBes})_{mn}(t) &= \sum_{k=0}^\infty J_{m+k}(t)J_{n+k}(t) \\
 &= \frac{t}{2(m-n)}\left[J_{m-1}(t)J_n(t)-J_m(t)J_{n-1}(t)\right],\quad m,n\in\mathbb{Z}.
 \label{eq:DBkernel}
\end{split}
\end{equation}
The discrete  Bessel kernel also appears in the studies of the polynuclear growth (PNG) model (e.g. \cite{prahofer2002scale}) and the longest increasing subsequence (LIS) in random permutation \cite{romik2015surprising}.
\noindent
The determinant that appears in eq. (\ref{eq:mgfwithC}) or eq. (\ref{eq:FDDBes}) is the Fredholm determinant, defined as
\begin{equation}
\mathrm{det}\left[1-zK\right]_{\ell^2(\mathbb{Z}_{>\alpha})}=1+\sum_{n=1}^\infty\frac{(-1)^n}{n!}z^n\sum_{x_1,\dots,x_n=\alpha+1}^\infty\mathrm{det}\left[K(x_j,x_k)\right]_{j,k=1,\dots,n}.
\label{eq:defofFD}
\end{equation}

By comparing the Taylor expansion at the point $z=1$ of this Fredholm determinant assuming $z=1-e^\lambda$, with the identity in eq. (\ref{eq:mgfexp}), we can see that
\begin{equation}
\mathrm{P}[N(t)=n]=\left.\frac{(-1)^n}{n!}\frac{d^n}{dz^n}\mathrm{det}\left[1-zK_\mathrm{DBes}(t)\right]\right|_{z=1}.
\label{eq:FCStoRMT}
\end{equation}
In \cite{eisler2013full}, it is also shown that the right hand side of eq. (\ref{eq:FCStoRMT}) has a meaning that exactly $n$ particles lie in the interval $m>\alpha$ and the analogous structures between FCS and Random Matrix Theory (RMT) was pointed out. For general instruction about RMT, see for instance \cite{mehta2004random,forrester2010log}.

Now an important observation for our analysis in this paper is that the Fredholm determinant with the discrete Bessel kernel in eq. (\ref{eq:FDDBes}) can also be written as the Fredholm determinant with the celebrated Bessel kernel \cite{forrester1993spectrum} for any parameter $z\in\mathbb{C}$,
\begin{equation}
\mathrm{det}\left[1-zK_\mathrm{DBes}(t)\right]_{\ell^2(\mathbb{Z}_{>\alpha})}=\mathrm{det}\left[1-zK_\mathrm{Bes}^{(\alpha)}\right]_{L^2[0,t^2]},
\label{eq:DBtoBes}
\end{equation}
where
\begin{equation}
K_\mathrm{Bes}^{(\alpha)}(x,y)=\frac{J_\alpha(\sqrt{x})\sqrt{y}J'_\alpha(\sqrt{y})-J'_\alpha(\sqrt{x})\sqrt{x}J_\alpha(\sqrt{y})}{2(x-y)}.
\label{eq:Bkernel}
\end{equation}
Our proof is based on a direct calculation of the trace which appears in the expansion of the logarithm of the determinants, see Appendix \ref{sec:AppendixA}.

The relation in eq. (\ref{eq:DBtoBes}) casts a fermion counting problem in the infinite and discrete spatial interval with fixed $t$ into the same problem but in the finite and continuous interval.
By considering $[0,t^2]$ as if it were a spatial interval, the latter situation can be taken as the non-interacting fermions in a confining potential and there have been considerable works on the subject (for example, see  \cite{dean2019noninteracting}).
Let $E(n;[0,t^2])$ denote the probability that $n$ fermions exist in the interval $[0,t^2]$ and write it with the Bessel kernel,
\begin{equation}
E(n;[0,t^2])=\left.\frac{(-1)^n}{n!}\frac{d^n}{dz^n}\mathrm{det}\left[1-zK_\mathrm{Bes}^{(\alpha)}\right]\right|_{z=1}.
\label{eq:E(n,J)}
\end{equation}
By observing that the expansion of both sides of eq. (\ref{eq:DBtoBes}) in terms of $z$ at $z=1$ are given by eq. (\ref{eq:FCStoRMT}) and eq. (\ref{eq:E(n,J)}), we have
\begin{equation}
\mathrm{P}[N(t)=n]=E(n;[0,t^2]).
\label{eq:P=E}
\end{equation}
In fact, we will study the large deviation of the left hand side of eq. (\ref{eq:P=E}) by analyzing that of the right hand side of eq. (\ref{eq:P=E}).

When $\alpha=0$, from eq. (\ref{eq:FCStoRMT}) we readily notice that the special case $n=0$ in eq. (\ref{eq:FCStoRMT}), is nothing but the return probability $R(t)$ (the quantum Loschmidt echo), which have been studied in \cite{prahofer2002scale,stephan2011local,viti2016inhomogeneous,allegra2016inhomogeneous,stephan2017return,krapivsky2018quantum}. It is defined as the modulus square of the overlap of the initial state and the final state, which is taken as the same as the initial state :
\begin{equation}
R(t):=|\braket{\mathrm{DW}|e^{-itH_{XX}}|\mathrm{DW}}|^2.
\end{equation}
For the case of domain wall initial state, and when we observe the up spins in the whole right subsystem $m>\alpha=0$ at time $t$, we have
\begin{equation}
R(t)=\mathrm{P}[N(t)=0]=\lim_{\lambda\to-\infty}\chi(\lambda,t)=\mathrm{det}\left[1-K_\mathrm{Bes}^{(0)}(x,y)\right]_{L^2[0,t^2]}.
\end{equation}
All representations here are totally equivalent. The meaning of the first equality is clear.
When the final state at time $t$ is completely the same as the initial state, there should be no particle in the right half of the chain $m>\alpha=0$.

The rightmost expression for the return probability as the Fredholm determinant is a special case ($z=1$) of the eq. (\ref{eq:DBtoBes})\footnote{The $z=1$ case can also be found by combining (8.96) in \cite{forrester2010log} and (4.3) in \cite{borodin2000fredholm}, as pointed out by P. Forrester.} and seems to have close connection with the expression appearing in \cite{stephan2017return} by considering the calculation of the partition function using  transfer matrix formalism in Euclidian time and some well-known determinant formula related to six-vertex model in two dimension with domain wall boundary conditions.
But we emphasize that the relation in eq. (\ref{eq:DBtoBes}), which holds for arbitrary $z$, and whose direct proof is given in Appendix \ref{sec:AppendixA}, will be crucial in our derivation of the large deviation function.

Actually, in the case of the XX spin chain with domain wall initial condition, a simple result for the return probability was obtained in $\cite{viti2016inhomogeneous,wei2017random}$
\begin{equation}
R(t)=e^{-\frac{t^2}{4}}
\label{eq:RP}
\end{equation}
for any time $t$. By taking $t$ large, we can calculate $\Psi(a=0)=1/4$.
This also makes us expect that the tail of the distribution of the integrated spin current decays as Gaussian as time passes as we will see later.

We can also evaluate the gradient at the leftmost edge of the large deviation tail.
Let us focus on the symmetric case $\alpha=0$ and write the following asymptotic formula for $r(n,t^2)$ which has been known for fixed $n$ and large $t$, see eq. (1.28) in \cite{tracy1994levelB},
\begin{equation}
E(n;[0,t^2])=E(0;[0,t^2])r(n;t^2)=e^{-t^2/4}r(n;t^2),
\end{equation}
\begin{equation}
r(n;t^2)\simeq G\left(n+1\right)^2\pi^{-n}2^{-n(2n+1)}t^{-n^2}e^{2nt},
\label{eq:r(n;J)}
\end{equation}
where $G(x)$ is the Barnes' G-Function \cite{olver2010nist}.
We now consider the large $n$ behavior of this.
With the help of the asymptotic expansion of the Barnes' G-Function $G(x)$ for large argument $z$ \cite{olver2010nist} (see chapter 5.17),
\begin{equation}
\log G\left(z+1\right)=\frac{1}{4}z^{2}+z\log\Gamma\left(z+1\right)-\left[\frac{1}{2}z(z+1)+\frac{1}{12}\right]\log z-\log A+\mathcal{O}\left(\frac{1}{z}\right),
\end{equation}
where $A=1.282427\dots$ is the Glaisher-Kinkelin constant, and using Stirling's approximation, we find
\begin{equation}
\begin{split}
E(n;[0,t^2]) &\simeq \exp\left[-\frac{t^2}{4}+\frac{1}{6}-2\log A-n\log2\pi+\left(n^2-\frac{1}{6}\right)\log n\right. \\
 &\quad \left.-\frac{3}{2}n^2-n\log\pi-(2n^2+n)\log2-n^2\log t+2nt\right]
\end{split}
\end{equation}
for large $t$ and $n$, but with the condition $n/t\ll 1$, because we considered the large $n$ limit after taking large $t$ limit.
When we scale the variable $n$ as $at$ due to the fact that mean current $\braket{N(t)}\simeq t/\pi$ is ballistic, we obtain
\begin{equation}
\mathrm{P}[N(t)=at,0\le a\ll1/\pi]\sim e^{-t^2\left(1/4-2a+3a^2/2-a^2\log a-a^2\log4\right)}
\end{equation}
for large $t$. From that asymptotics, we can read that the gradient near the origin would be given as $\Psi'(0)=-2$.

We remark that, even though our main focus of this work is for a finite $\alpha$, in particular the $\alpha=0$ case, some formulas in this paper should also be useful for studying the fluctuation of the quantum front (by setting the position of  measuring current as $\alpha=t+2^{-1/3}t^{1/3}s$) for large $t$ \cite{hunyadi2004dynamic,eisler2013full,viti2016inhomogeneous,stephan2019free}.
Here $s$ is a local coordinate describing the wavefront.
In \cite{borodin2003increasing}, it was known that the regime described by the Bessel kernel with the interval $[0,Q_\alpha(s)]$, where $Q_\alpha(s)=(\alpha-2^{-1/3}\alpha^{1/3}s)^2$, tends to be that described by the Airy kernel \cite{forrester1993spectrum} with the interval $[s,\infty)$, in the large $\alpha$ limit.
By identifying $Q_\alpha(s)$ as $t^2$ and considering large $\alpha$ limit, we can focus on the statistics of up spins in the front regime.

When we evaluate the asymptotic behavior of the number variance of integrated current $N(t)$ in Appendix \ref{sec:AppendixC}, it is useful to work with the variable $\sqrt{x}\rightarrow x$ in eq. (\ref{eq:Bkernel}), which leads to a transformed Bessel kernel,
\begin{equation}
\tilde{K}_\mathrm{Bes}^{(\alpha)}(x,y)=\sqrt{xy}\frac{J_\alpha(x)yJ'_\alpha(y)-J'_\alpha(x)xJ_\alpha(y)}{x^2-y^2}.
\label{eq:sBkernel}
\end{equation}
In this case, the kernel $\tilde{K}_\mathrm{Bes}^{(\alpha)}$ in eq. (\ref{eq:sBkernel}) acts on $L^2[0,t]$.
Thus, expanding the Fredholm determinant with regard to $\tilde{K}_\mathrm{Bes}^{(\alpha)}$ with the use of a continuous version of eq. (\ref{eq:defofFD}), we have
\begin{equation}
\chi(\lambda,t)=1+\sum_{n=1}^\infty\frac{1}{n!}(e^\lambda-1)^n\int_0^t\prod_{k=1}^ndt_k\mathrm{det}\left[\tilde{K}_\mathrm{Bes}^{(\alpha)}(t_j,t_k)\right]_{j,k=1,\dots,n}.
\end{equation}

\section{Bessel kernel in RMT}

Here, we briefly review how the Bessel kernel emerges in RMT.
Let $X$ be an $M\times N~(M\ge N)$ Gaussian matrix whose elements are complex and follow the Gaussian distribution having unit variance but zero mean. Then, the $N\times N$ matrix
\begin{equation}
W=X^\dagger X
\end{equation}
is called the Wishart matrix which is originally from \cite{wishart1928generalised}. The joint eigenvalue probability density function of the Wishart matrix is given by \cite{james1964distributions}
\begin{equation}
P(\{x_i\}_{i=1}^N)\propto\prod_{i=1}^N x_i^{\alpha}e^{-x_i}\prod_{1\le j<k\le N}|x_j-x_k|^2,
\label{eq:jpdf}
\end{equation}
where $\alpha=M-N$.
Note that all the eigenvalues are positive $x_i\ge 0$.

If we see the eigenvalues as the positions of identical fermionic particles, the joint eigenvalue probability density function can be represented by $N$ point function of fermions as
\begin{equation}
\prod_{i=1}^N x_i^{\alpha}e^{-x_i}\prod_{1\le j<k\le N}|x_j-x_k|^2\propto\mathrm{det}\left[K_N(x_j,x_k)\right]_{j,k=1,\dots,N},
\end{equation}
where the kernel $K_N(x,y)$ can be constructed in terms of wave functions $\{\phi_k(x)\}_{k=0}^{N-1}$,
\begin{equation}
K_N(x,y)=\sum_{k=0}^{N-1}\phi_k(x)\phi_k(y).
\end{equation}
Since the weight function for this case is $x^{\alpha}e^{-x}$, the wave function can be written in terms of the generalized Laguerre polynomials $L_{k}^{(\alpha)}(x)$ as
\begin{equation}
\phi_k(x)=\sqrt{\frac{k!}{\Gamma(k+\alpha+1)}}x^{\alpha/2}e^{-x/2}L_{k}^{(\alpha)}(x).
\end{equation}
With the help of the Christoffel Darboux formula \cite{szego1939orthogonal}, we have
\begin{equation}
K_N(x,y)=\frac{N!}{\Gamma(N+\alpha)}\sqrt{\frac{\Gamma(N+\alpha+1)\Gamma(N+\alpha)}{N!(N-1)!}}\frac{\phi_{N-1}(x)\phi_N(y)-\phi_N(x)\phi_{N-1}(y)}{x-y}.
\end{equation}
It is worth checking that the one point function after a proper scaling \cite{bronk1965exponential} gives the Mar\v{c}enko and Pastur distribution \cite{marvcenko1967distribution}
\begin{equation}
\lim_{N\to \infty}K_N\left(Nx,Nx\right)=\frac{1}{2\pi}\sqrt{\frac{4-x}{x}}.
\end{equation}

When we study behaviors of eigenvalues near the hard edge at $x\simeq0$, the Bessel kernel appears after the scaling \cite{forrester1993spectrum},
\begin{equation}
\lim_{N\to \infty}\frac{1}{4N}K_N\left(\frac{x}{4N},\frac{y}{4N}\right)=K_\mathrm{Bes}^{(\alpha)}(x,y).
\label{eq:lim2ptfn}
\end{equation}
In order to get this kernel, the following large $N$ asymptotics formula of the generalized Laguerre polynomials $L_{k}^{(\alpha)}(x)$ \cite{szego1939orthogonal} is helpful,
\begin{equation}
x^{\alpha/2}e^{-x/2}L_{N}^{(\alpha)}(x)\simeq N'^{\alpha/2}\frac{\Gamma(N+\alpha+1)}{N!}J_\alpha(2\sqrt{N'x}).
\end{equation}
Here, a shorthand notation $N'=N+(\alpha+1)/2$ is used.

Thus, we can see that the Bessel kernel describes the statistics of eigenvalues at the hard edge where the smallest eigenvalue of $N\times N$ Wishart matrix is located.
To evaluate the probability measure $E(n;[0,t^2])$ in an exponential form, we consider the corresponding probability measure $E_N(n;[0,t^2/4N])$ that $n$ eigenvalues lie in the interval $[0,t^2/4N]$ for a large but finite $N$ Wishart ensemble.
The change of the interval as compared to eq. (\ref{eq:E(n,J)}) follows from the scaling of the argument appearing in eq. (\ref{eq:lim2ptfn}).

For large $N$ with finite and fixed $\alpha$, the contribution from the one body potential $\alpha\log x$ can be ignored compared to another one body potential $x/2$.
Therefore, we only consider the case $\alpha=0$ from now on.
Going back to the joint eigenvalue probability density function in eq. (\ref{eq:jpdf}) with $\alpha=0$ of  the Wishart matrix, let us write this probability in an exponential form
\begin{equation}
P(\{x_i\}_{i=1}^N)\propto\prod_{i=1}^N e^{-x_i}\prod_{1\le j<k\le N}|x_j-x_k|^2=e^{-2U(\{x_i\}_{i=1}^N)}.
\end{equation}
This can be interpreted as the Boltzmann factor and the exponent plays a role of the total energy of the $N$ particle system
\begin{equation}
U(\{x_i\}_{i=1}^N)=\sum_{i=1}^N\frac{x_i}{2}-\sum_{j<k}\log|x_j-x_k|.
\label{eq:Energy}
\end{equation}
The second term implies that this energy is the energy of gases having two dimensional Coulomb interaction as an analogy.

\section{Number statistics via coulomb gas method}

Because of the analogy mentioned at the end of the previous section, the Coulomb gas method \cite{dyson1962statistical,forrester2010log} is widely used to determine certain properties of fermions confined in some potential at zero temperature (see, e.g. \cite{cunden2016shortcut} and references therein).
In  \cite{majumdar2012number,marino2016number}, large deviation properties of the Wishart matrix were studied with the method, 
with the normalization such that a scaling $x\rightarrow Nx$ is performed in eq. (\ref{eq:jpdf}). 
The authors in \cite{majumdar2012number,marino2016number} showed that 
the probability $E_N(n;I)$, that there are $n$ eigenvalues in an arbitrary interval $I (\subseteq[0,\infty))$  for $N\times N$ 
Wishart matrix with this normalization, satisfies the following large deviation property in the large $N$ limit,
\begin{equation}
E_N(n=\kappa N;I)\sim e^{-2N^2\psi_I(\kappa)},\quad\mathrm{for}\quad0\le\kappa\le1.
\label{LDPWis}
\end{equation}
We will utilize their result to our problem. Notice that, with the change of normalization 
of the Wishart matrix mentioned above, the discussions in the previous section tells us that the probability (\ref{eq:E(n,J)}),  
associated with the Bessel kernel, is recovered in the $N\to\infty$ limit as 
\begin{equation}
\lim_{N\to\infty}E_N\left(n;\left[0,\frac{t^2}{4N^2}\right]\right)=E(n,[0,t^2]).
\label{ENlim}
\end{equation}

Now we briefly review description of the large deviation function $\psi_I$ from  
\cite{majumdar2012number,marino2016number}, for the case of 
the interval $I=[0,t^2/4N^2]$ of our interest. 
Given the interval, and the number of eigenvalues or the fermions $n$ in the interval $I$, the most probable distribution of the spectrum density $\rho^\ast(x)$ is determined as a result of minimization problem of the free energy under the constraint.
Here the density $\rho^\ast(x)$ should be taken so that it satisfies the normalization condition :
\begin{equation}
\int_0^\infty\rho^\ast(x)dx=1.
\label{eq:Normalizedrho}
\end{equation}

Since $\rho^\ast(x)$ which minimizes the total energy in the Boltzmann factor corresponding to the expression in (\ref{eq:jpdf}) gives the main contribution to the probability distribution, finding such a density $\rho^\ast(x)$ is a crucial problem.
Indeed, $\rho^\ast(x)$ exists and the following statement is known.
The large deviation function $\psi_I(\kappa)$ for large $N$ is given as \cite{majumdar2012number,marino2016number}
\begin{equation}
\psi_I(\kappa)=\frac{1}{2}\int_0^\infty \frac{x}{2}\rho^\ast(x)dx-\frac{\mu}{2}\kappa-\frac{\eta}{2}-\frac{3}{4}.
\label{eq:LDFforN}
\end{equation}
Here $\mu$ and $\eta$ in our case are given by
\begin{align}
\mu &= -\int_{\min\{\lambda_-,t^2/4N^2\}}^{\max\{\lambda_-,t^2/4N^2\}} \left[G(x)-\frac{1}{2}\right]dx, \\
\eta &= \log{\lambda_+}-\frac{\lambda_+}{2}-\int_{\lambda_+}^\infty\left[G(x)-\frac{1}{x}\right]dx,
\end{align}
where $\lambda_\pm$ denotes the two spectral edges of the four, as explained later.
In fact, the supports of $\rho^\ast(x)$ will be found to be
\begin{equation}
\mathrm{supp}(\rho^\ast)(x)=
\begin{cases}
\quad(0,\lambda_-]\cup(t^2/4N^2,\lambda_+], &\mathrm{for}\quad 0\le n<t/\pi, \\[2ex]
\quad(0,t^2/4N^2)\cup[\lambda_-,\lambda_+], &\mathrm{for}\quad t/\pi<n.
\end{cases}
\end{equation}
Additionally, the resolvent $G(x)$ is defined by the Stieltjes transformation of the density $\rho^\ast(x)$ as
\begin{equation}
G(z)=\int\frac{\rho^\ast(x)}{z-x}dx.
\label{eq:defG(z)}
\end{equation}

In fact, $G(x)$ satisfies the self-consistent and quadratic equation coming from the saddle point equation with regards to $\rho^\ast(x)$.
The two solutions are given as
\begin{equation}
G_\pm(z)=\frac{1}{2}\pm\frac{1}{2}\sqrt{\frac{(z-\lambda_+)(z-\lambda_-)}{z(z-t^2/4N^2)}}.
\label{eq:G_pm(z)}
\end{equation}
The sign between the first and the second terms is to be determined by the density $\rho^\ast(x)$.
Explicitly, it is given in eq. (\ref{eq:Gr<1}) and in eq. (\ref{eq:Gr>1}).
In this resolvent, $\lambda_+>\lambda_-$ are the roots of the numerator of the fraction inside the square root.
From the definition of the resolvent in eq. (\ref{eq:defG(z)}), it should decay like $G(z)\simeq1/z$ for large $z$ as the density $\rho^\ast(x)$ is normalized as in eq. (\ref{eq:Normalizedrho}).
Therefore, by eq. (\ref{eq:G_pm(z)}), we have
\begin{equation}
\lambda_++\lambda_--\frac{t^2}{4N^2}=4.
\label{eq:propofG}
\end{equation}

We can extract the spectral density $\rho^\ast(x)$ from the imaginary part of this resolvent as
\begin{equation}
\rho^\ast(z)=-\frac{1}{\pi}\lim_{\varepsilon\to+0}\mathrm{Im}G(z+i\varepsilon)=\frac{1}{2\pi}\sqrt{\frac{(\lambda_+-z)(\lambda_--z)}{z(t^2/4N^2-z)}}.
\label{eq:mean field}
\end{equation}
Now we can identify the role of parameters $\lambda_\pm$, and $t^2/4N^2$.
The spectral density $\rho^\ast(x)$ has two compact supports with four edges.
The leftmost endpoint $x=0$ and the rightmost endpoint $x=\lambda_+$ are always unchanged for large $N$, while the order of the rest endpoints $x=\lambda_-$ and $x=t^2/4N^2$ could change depending on the value of $\lambda_-$.

The condition that this spectral density contains $n$ eigenvalues in the interval $I=[0,t^2/4N^2]$ is reflected in another constraint,
\begin{equation}
\int_I\rho^\ast(x)dx=\kappa=\frac{n}{N}.
\label{eq:numofevinI}
\end{equation}

The two constraints in eq. (\ref{eq:propofG}) and eq. (\ref{eq:numofevinI}) determine the two parameters $\lambda_\pm$ uniquely as functions of $n$, $t$ and $N$.
Consequently these constraints implies that when $n<t/\pi$, we can see $\lambda_-<t^2/4N^2$ while when $n>t/\pi$, we can see $t^2/4N^2<\lambda_-$.
In order to focus on to the hard edge regime $x\simeq0$, we need to take the large $N$ limit as explained in the previous section.

Before we tackle the large deviation, let us see the mean value $\braket{N(t)}$ with eq. (\ref{eq:numofevinI}).
For large $N$, the interval $I$ around $x\simeq0$ having the order $\mathcal{O}(1/N^2)$ shrinks to zero and the Mar\v{c}enko and Pastur distribution \cite{marvcenko1967distribution} in $I$ diverges as $\mathcal{O}(N)$ as $N$ increases.
Since the spectral density contains the total number of eigenvalues $N$ inside its support, multiplying $\rho^\ast(x)$ by $N$, we can extract a meaningful and finite quantity.
It turns out to be 
\begin{equation}
\lim_{N\to\infty}\frac{N}{2\pi}\int_0^{t^2/4N^2}\sqrt{\frac{4-x}{x}}dx=\frac{t}{\pi}.
\end{equation}
The result is consistent with the previous result shown in eq. (\ref{eq:AveandVar}).

Since we are considering the deviation from the mean value $\braket{N(t)}\simeq t/\pi$, the number of transferred up spins $n$ should have the same order as the mean.
Thus, it is reasonable to expect that the parameter $\lambda_-$ which adjusts $n$ and $t^2/4N^2$ have the same order.
For this reason, we could introduce the ratio $r$ that satisfies
\begin{equation}
\lambda_-=\frac{t^2}{4N^2}r.
\label{defofr}
\end{equation}
When $r=1$, the spectral distribution $\rho^\ast$ becomes the Mar\v{c}enko and Pastur distribution \cite{marvcenko1967distribution}, since $\lambda_+=4$ and $\lambda_-=t^2/4N^2$.

Next, we will analyze the large deviation.
As a preparation, we first expand the large deviation function $\psi_I(\kappa)$ as a series in $1/N$ with $t$ fixed.
Details of the calculation will be given in Appendix \ref{sec:AppendixB}. 
As a result of expansion, the terms having the order $\mathcal{O}(1)$ and  $\mathcal{O}(1/N)$ do not remain.
We find
\begin{equation}
\psi_I(n/N)=\frac{t^2}{8N^2}(1-r)+\frac{n}{2N^2}t\sqrt{r}\left[K\left(1-\frac{1}{r}\right)-E\left(1-\frac{1}{r}\right)\right]+\mathcal{O}\left(\frac{1}{N^3}\right)
\label{psiIexp}
\end{equation}
where $K(k^2)$ and $E(k^2)$ are the complete elliptic integrals of the first and second kind \cite{olver2010nist} respectively. The definitions are as follows
\begin{equation}
\begin{split}
K(k^2) &= \int_0^{\pi/2}\frac{d\theta}{\sqrt{1-k^2\sin^2{\theta}}}, \\
E(k^2) &= \int_0^{\pi/2}\sqrt{1-k^2\sin^2{\theta}}d\theta.
\end{split}
\label{eq:KandE}
\end{equation}
By definition, $K(0)=E(0)=\pi/2$ follows immediately.
Furthermore, we can also check the asymptotic relation such as $K(-k^2)\simeq(\log4k)/k$ and $E(-k^2)\simeq k$ for large $k>0$ from \cite{olver2010nist,friedman1954handbook}.
The subleading terms $\mathcal{O}(1/N^3)$ are not important because those should be discarded after the limit $N\rightarrow\infty$.

If we set $n=at$ with a new parameter $a$ in (\ref{psiIexp}), the leading order terms are proportional to $t^2/N^2$. 
Multiplying $2N^2$ in front of the large deviation function $\psi_I(\kappa)$ as we have seen in (\ref{LDPWis}) for large $N$, we 
get terms proportional to $t^2$ in the exponent.
Though the expansion in $1/N$ in Appendix \ref{sec:AppendixB} is performed for a fixed $t$,
one should consider large $t$ limit because the large deviation results in \cite{majumdar2012number,marino2016number} were obtained for large $n(=at)$. 
Namely, we should consider the limits of large $t$ and $N$ but with the condition $t\ll N$. 
In this limit, one can expect that, although there could be some subleading terms, 
the large deviation behaviors for large $t$ are captured by the terms proportional to $t^2$ which were found above. 

\begin{figure}
\begin{tabular}{c}
\begin{minipage}[b]{0.5\hsize}
\centering
\includegraphics[width=60mm,clip]{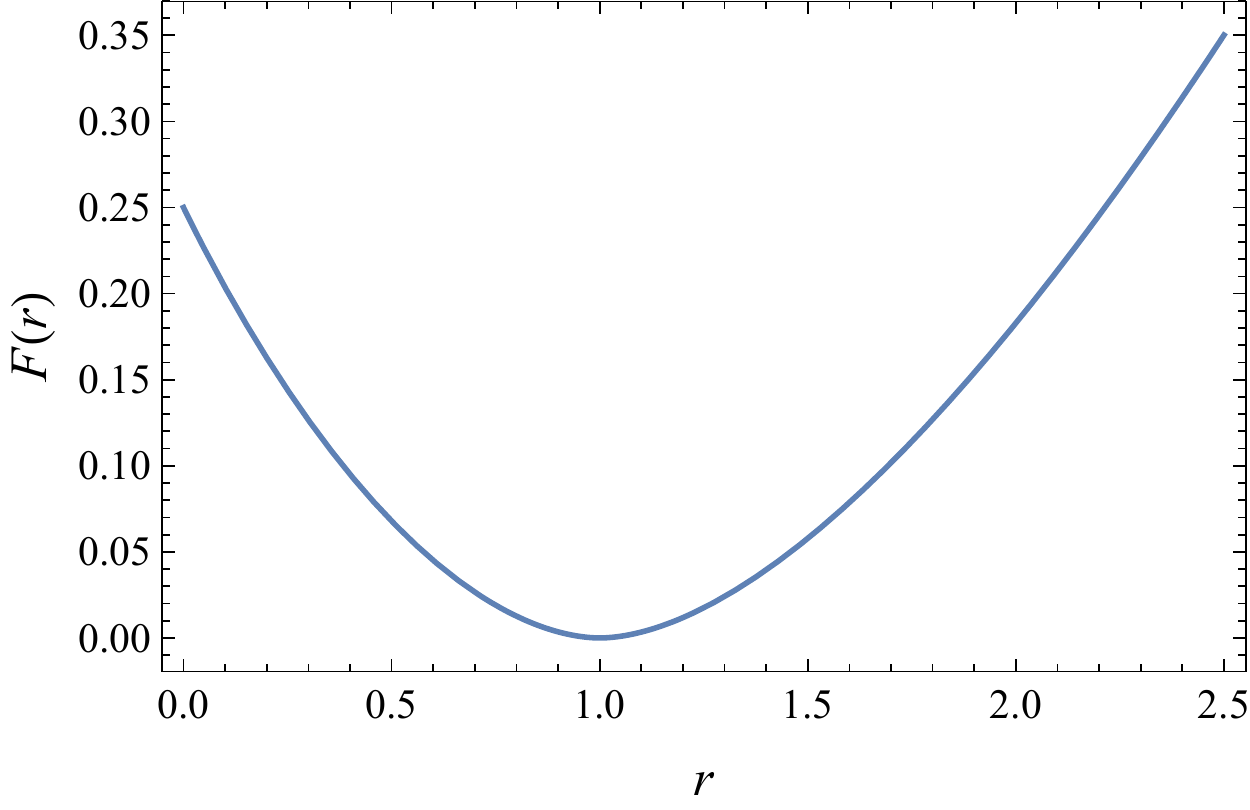}
\vspace{-0.5pt}
\caption{The graph of the function $F(r)$ versus $r$.}
\vspace{-2pt}
\label{fig:F_vs_r}
\end{minipage}
\hspace{10pt}
\begin{minipage}[b]{0.5\hsize}
\centering
\includegraphics[width=60mm,clip]{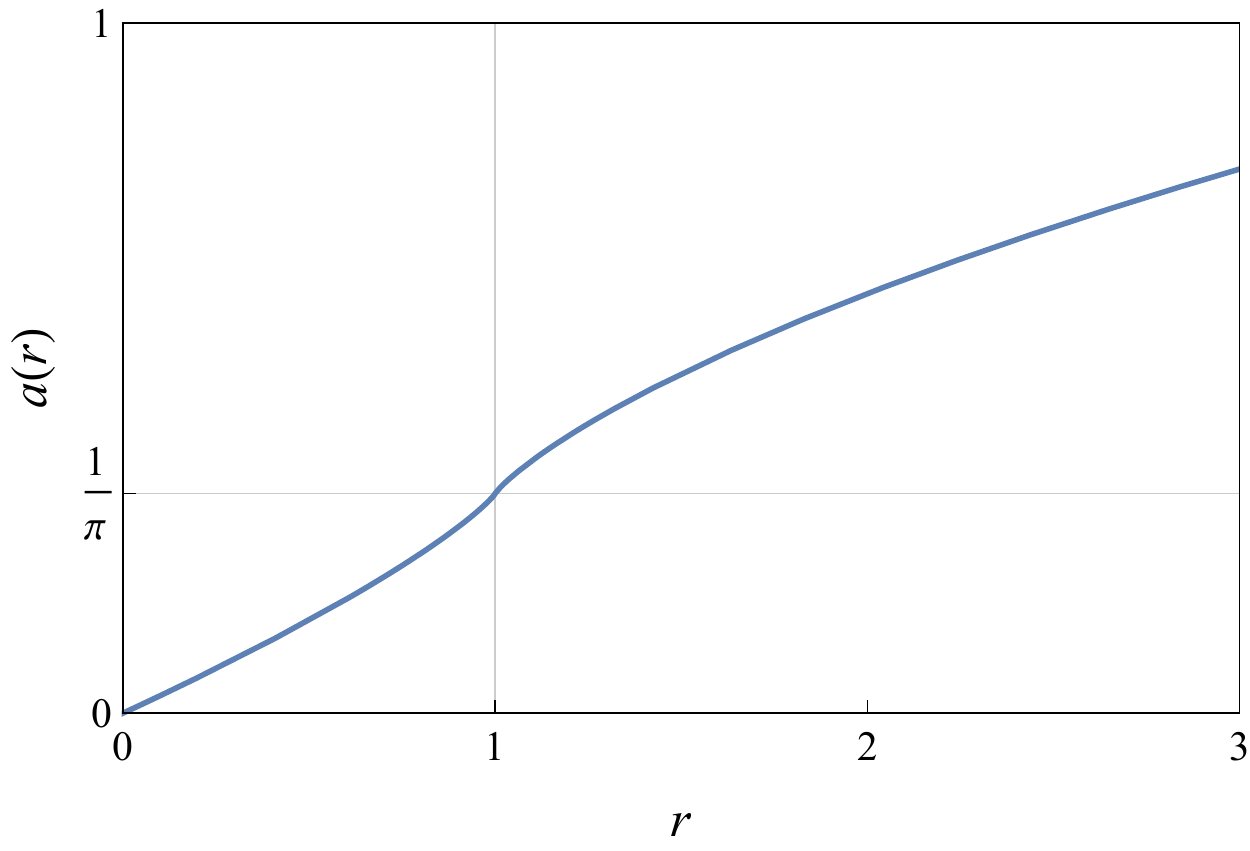}
\caption{The relation between $a$ and $r$.}
\label{fig:a_vs_r}
\end{minipage}
\end{tabular}
\end{figure}

Recalling eqs. (\ref{eq:P=E}) and (\ref{ENlim}), and substituting (\ref{psiIexp}) into (\ref{LDPWis}), we finally obtain
\begin{equation}
\mathrm{P}[N(t)=at]=\lim_{N\to\infty}E_N\left(at;\left[0,\frac{t^2}{4N^2}\right]\right)\sim e^{-t^2\Psi(a)},
\label{eq:Nlimit}
\end{equation}
where the desired large deviation function $\Psi(a)=F(r(a))$ is found as
\begin{equation}
F(r)=\frac{1}{4}(1-r)+a(r)\sqrt{r}\left[K\left(1-\frac{1}{r}\right)-E\left(1-\frac{1}{r}\right)\right],
\label{eq:LDF}
\end{equation}
with the following condition  from eq. (\ref{eq:a(r)<t/pi}) and eq. (\ref{eq:a(r)>t/pi}) :
\begin{equation}
a(r)=
\begin{cases}
\quad\dfrac{1}{\pi}\sqrt{1-r}\left[E\left(\dfrac{r}{r-1}\right)-K\left(\dfrac{r}{r-1}\right)\right], &\mathrm{for}\quad 0\le r<1, \\[4ex]
\quad\dfrac{1}{\pi}\sqrt{r}E\left(\dfrac{1}{r}\right), &\mathrm{for}\quad 1<r,
\end{cases}
\label{eq:ator}
\end{equation}
which is derived by considering large $N$ behaviors of the integral in eq. (\ref{eq:numofevinI}), see Appendix \ref{sec:AppendixB}.
The shapes of $F(r)$ and $a(r)$ are shown in Fig. \ref{fig:F_vs_r} and Fig. \ref{fig:a_vs_r}.
The large deviation function $\Psi(a)$  is also shown as the solid curve in Fig. \ref{fig:LDF} below 
(together with the result by DMRG explained in section \ref{sec:dmrg}). 
The graph is slightly asymmetric with respect to the refection at $a=1/\pi$.

Restoring the ferromagnetic coupling constant in eq. (\ref{eq:Nlimit}), we obtain eq. (\ref{eq:IntroResult}),
which is our main result in the paper. 

As one sees in Fig. \ref{fig:a_vs_r},  as $a$ increases, $r$ also tends to increase.
When $a$ takes a values from $0$ to $1/\pi$, the parameter $r$ varies from $0$ to $1$ and when $a$ takes a value more than $1/\pi$, $r$ also varies from $1$ to $\infty$.
The function $a(r)$  is always invertible as $r(a)$.

We can check that the minimum of $\Psi(a)$ is attained at $a=1/\pi$ and this value is zero.
First, the function $F(r)$ has indeed the minimum at $r=1$.
Using the chain rule, the derivative of $\Psi(a)$ with regard to $a$ is
\begin{equation}
\begin{split}
\frac{d\Psi(a)}{da} &= \frac{dF(r)}{dr}\frac{dr}{da} \\
 &= -\frac{1}{4a'(r)}+\sqrt{r}\left[K\left(1-\frac{1}{r}\right)-E\left(1-\frac{1}{r}\right)\right]+\frac{a(r)}{2\sqrt{r}a'(r)}K\left(1-\frac{1}{r}\right).
\end{split}
\label{eq:DPsi(a)}
\end{equation}
Here, the derivative of $a$ with regards to $r$ is given by
\begin{equation}
\frac{da(r)}{dr}=
\begin{cases}
\quad\dfrac{1}{2\pi\sqrt{1-r}}K\left(\dfrac{r}{r-1}\right), &\mathrm{for}\quad 0\le r<1, \\[4ex]
\quad\dfrac{1}{2\pi\sqrt{r}}K\left(\dfrac{1}{r}\right), &\mathrm{for}\quad 1<r.
\end{cases}
\label{eq:Da}
\end{equation}
This derivative takes a positive value for all $r>0$.
Therefore, the sign of the sum of the first and the third term in eq. (\ref{eq:DPsi(a)}) should be determined by the following terms,
\begin{equation}
\frac{a(r)}{2\sqrt{r}}K\left(1-\frac{1}{r}\right)-\frac{1}{4}.
\end{equation}
Substituting $a(r)$ in eq. (\ref{eq:ator}) tells us that the terms above are monotonically increasing function in terms of $r$.
Noticing that
\begin{equation}
\lim_{r\to1}\frac{a(r)}{2\sqrt{r}}K\left(1-\frac{1}{r}\right)=\frac{1}{4},
\end{equation}
the sum of the first and third term in eq. (\ref{eq:DPsi(a)}) is negative for $0\le r<1$ and positive for $r>1$.
We can also confirm that the second term in eq. (\ref{eq:DPsi(a)}) takes negative value for $0\le r<1$ and positive value for $r>1$ from the behavior of the elliptic integrals $E$ and $K$.
Hence, the large deviation function $F(r)$ has an unique global minimum at $r=1$.
Recalling that $r=1$ corresponds to $a=1/\pi$ from eq. (\ref{eq:ator}), we can confirm that
\begin{equation}
\Psi\left(a=\frac{1}{\pi}\right)=0.
\end{equation}
The Gaussian fluctuation around $a=1/\pi$, will be studied in the next section.

In section \ref{sec:FCSandRMT}, we estimated that the large deviation function should behave as $\Psi(0)=1/4$ and $\Psi'(0)=-2$ at the leftmost endpoint.
Now, we can also check the correctness of the estimates by using the explicit representation of large deviation function derived above.
In evaluating those, we can utilize the asymptotic behavior $K(-k^2)\simeq(\log4k)/k$ and $E(-k^2)\simeq k$ for large $k>0$, mentioned above.
As shown in Fig. \ref{fig:a_vs_r}, $r$ obviously goes to $0$ as well as $a$.
Therefore we only consider $r\rightarrow0$ limit.

First, using $a(0)=0$ and $E(1-1/r)\simeq1/\sqrt{r}$ for small $r$, we can easily confirm that
\begin{equation}
\lim_{a\to0}\Psi(a)=\lim_{r\to0}F(r)=\frac{1}{4}.
\end{equation}
Here we used the fact that $K(1-1/r)$ goes to zero when taking the limit $r\rightarrow0$.

Second, we confirm that $\Psi'(0)=-2$.
The function $\Psi(a)$ has already differentiated in eq. (\ref{eq:DPsi(a)}).
Here the following two terms in eq. (\ref{eq:DPsi(a)}) vanish in the limit $r\rightarrow0$,
\begin{equation}
\lim_{r\to0}\sqrt{r}K\left(1-\frac{1}{r}\right)=\lim_{r\to0}\frac{a(r)}{2\sqrt{r}a'(r)}K\left(1-\frac{1}{r}\right)=0.
\end{equation}
Next, the derivative of $a$ with regards to $r$ for $0\le r<1$ in eq. (\ref{eq:Da}) gives $a'(0)=1/4$ and thus the first term in eq. (\ref{eq:DPsi(a)}) is equal to $-1$.
Considering the limit of the contribution from $-\sqrt{r}E(1-1/r)$ after using the asymptotics $E(-1/r)\simeq1/\sqrt{r}$, we can see that $-\sqrt{r}E(1-1/r)$ also gives $-1$.
Combining the above calculation, we obtain
\begin{equation}
\lim_{a\to0}\Psi'(a)=\lim_{r\to0}\frac{F'(r)}{a'(r)}=-2.
\end{equation}

\section{Gaussian fluctuation with variance of $\mathcal{O}\left(\log t\right)$}
\label{sec:Variance}

As already mentioned in the introduction, in \cite{antal2008logarithmic}, the variance was shown to behave as $\braket{N(t)^2}_c\simeq(\log{t}+C)/2\pi^2$ and the constant $C$ was written as a sum of a few mathematical constants and definite integrals.
This behavior is also observed in the context of the hard edge statistics in RMT and it is proven that the normalized random variable $(N(t)-\braket{N(t)})/\sqrt{\braket{N(t)^2}_c}$ converges to the Gaussian distribution \cite{soshnikov2000gaussian}.
The proof of the convergence is based on the discussion in \cite{costin1995gaussian}.
In this section, we show one can recover this behavior by expanding the large deviation function obtained in the previous section around the most probable point $a=1/\pi$.
As a byproduct, we find a much simpler formula for the constant $C$, see eq. (\ref{eq:Var}).
To see the main contribution, we can follow the same procedure as \cite{majumdar2012number,marino2016number}, and as in their case, we can see the property that the large deviation function at the minimum is not analytical inherit.
Therefore, we need to see the increment which emerges when we slide the value $a$ towards the horizontal axis only by small difference $\delta a>0$.
As the eq. (\ref{eq:ator}) gives the relation between $\delta a>0$ and $\delta r>0$, we can access the increment $\delta F$.

In the case of $0<a<1/\pi$ or $0<r<1$, we have
\begin{equation}
\begin{split}
F(r) &= \frac{1}{4}(1-r)+\dfrac{1}{\pi}\sqrt{r(1-r)} \\
 &\quad \times\left[E\left(\dfrac{r}{r-1}\right)-K\left(\dfrac{r}{r-1}\right)\right]\left[K\left(1-\frac{1}{r}\right)-E\left(1-\frac{1}{r}\right)\right].
\end{split}
\end{equation}
Decreasing the parameter by $\delta r>0$, expansion of the function $F(r)$ in terms of $\delta r$ is as follows
\begin{equation}
F(1-\delta r)=\frac{1}{32}\left(1+8\log2-2\log\delta r\right)(\delta r)^2+\mathcal{O}\left((\delta r)^3\log\delta r\right).
\label{eq:-dr}
\end{equation}
The transformation $a\rightarrow a-\delta a$ can be read from the eq. (\ref{eq:ator}) as
\begin{equation}
-\delta a=-\frac{1}{4\pi}\left(1+4\log2-\log\delta r\right)\delta r+\mathcal{O}\left((\delta r)^2\log\delta r\right).
\label{eq:ator-}
\end{equation}

In the case of $1/\pi<a$ or $1<r$, we have
\begin{equation}
F(r)=\frac{1}{4}(1-r)+\dfrac{1}{\pi}rE\left(\frac{1}{r}\right)\left[K\left(1-\frac{1}{r}\right)-E\left(1-\frac{1}{r}\right)\right].
\end{equation}
The same calculation as above gives
\begin{equation}
F(1+\delta r)=\frac{1}{32}\left(1+8\log2-2\log\delta r\right)(\delta r)^2+\mathcal{O}\left((\delta r)^3\log\delta r\right).
\label{eq:+dr}
\end{equation}
The transformation $a\rightarrow a+\delta a$ can be read from the eq. (\ref{eq:ator})
\begin{equation}
\delta a=\frac{1}{4\pi}\left(1+4\log2-\log\delta r\right)\delta r+\mathcal{O}\left((\delta r)^2\log\delta r\right).
\label{eq:ator+}
\end{equation}

The relation between $\delta a$ and $\delta r$ in eq. (\ref{eq:ator-}) is, in fact, the same as that in eq. (\ref{eq:ator+}).
That is why it is enough to consider one of them.
Inverting this relation, we have
\begin{equation}
\frac{\delta r}{16e}=\exp W\left(-\frac{\pi}{4e}\delta a\right)=\frac{-\frac{\pi}{4e}\delta a}{W(-\frac{\pi}{4e}\delta a)}
\label{eq:rtoa}
\end{equation}
where $W(x)$ is the Lambert W-function \cite{olver2010nist} (see chapter 4.13), which is defined to be the solution of the following equation
\begin{equation}
We^W=x.
\end{equation}
For $x<0$, notice that $W(x)$ has a branch because $W(x)$ is a multi-valued function, and we should take the one which takes the range of the function in $(\infty,-1/e]$ in the calculations below. For $W(x)\le-1/e$, this function has an expansion
\begin{equation}
W(x)=-\eta-\ln\eta-\frac{\ln\eta}{\eta}+\mathcal{O}\left(\frac{(\ln\eta)^2}{\eta^2}\right)
\end{equation}
as $x\rightarrow-0$ where $\eta=\ln\left(-1/x\right)$.

By using this asymptotic expansion of the Lambert W-function $W(x)$ and substituting eqs. (\ref{eq:ator-}), (\ref{eq:rtoa}) and eq. (\ref{eq:ator+}), (\ref{eq:rtoa}) into eq. (\ref{eq:-dr}) and eqs. (\ref{eq:+dr}) respectively, we obtain
\begin{equation}
F(1\pm\delta r)\simeq-\frac{1}{16}\left(\log\delta r-4\log2-1\right)(\delta r)^2+\mathcal{O}\left((\delta r)^2\right).
\end{equation}
Therefore, after we restore the scale of $\delta n=t\delta a $,
\begin{equation}
-t^2\Psi(1/\pi\pm\delta a)\simeq t^2e\delta r\delta a\simeq-\frac{1}{2}\frac{2\pi^2}{\log t+\mathcal{O}\left(1\right)}(\delta n)^2.
\label{eq:Gaussian}
\end{equation}
Inserting this into eq. (\ref{eq:Nlimit}), we could derive the main contribution to the variance, which increases logarithmically in time with the coefficient $1/2\pi^2$, as shown in \cite{antal2008logarithmic}.

To determine the number variance up to the correction $\mathcal{O}(1)$ in eq. (\ref{eq:Gaussian}), we need to carry out the direct calculation by using the Bessel kernel.
In Appendix \ref{sec:AppendixC}, we show
\begin{equation}
\braket{N(t)^2}_c=\frac{1}{2\pi^2}\left(\log t+\log4+\gamma+1\right),
\label{eq:Var}
\end{equation}
see eq. (\ref{eq:AppVar}).
Therefore we have found that the constant $C$ in eq. (\ref{eq:AveandVar}) is, in fact, simply given by
\begin{equation}
C=\log4+\gamma+1=2.9635100260\dots,
\end{equation}
which numerically matches precisely with the value from the calculation in \cite{antal2008logarithmic}.

In principle, we can address the cumulants higher than the second one by using the formula shown in eq. (\ref{eq:CumulantExp}) as we calculate the second cumulant in Appendix \ref{sec:AppendixC} though the full direct calculation of all the cumulants is cumbersome and remains as a future problem.
We can see numerically that a few subsequent cumulants are suppressed compared to the second cumulant and are around zero.
This observation also supports the fact that higher-order cumulants are smaller than the second cumulant which diverges logarithmically in time \cite{soshnikov2000gaussian} and, as a consequence, that the fluctuation of the spin current is the Gaussian when focusing on the vicinity of the mean on a scale $\mathcal{O}(\sqrt{\log t})$.

\section{Numerical Analysis}
\label{sec:dmrg}

\begin{figure}
\centering
\includegraphics[width=10cm,clip]{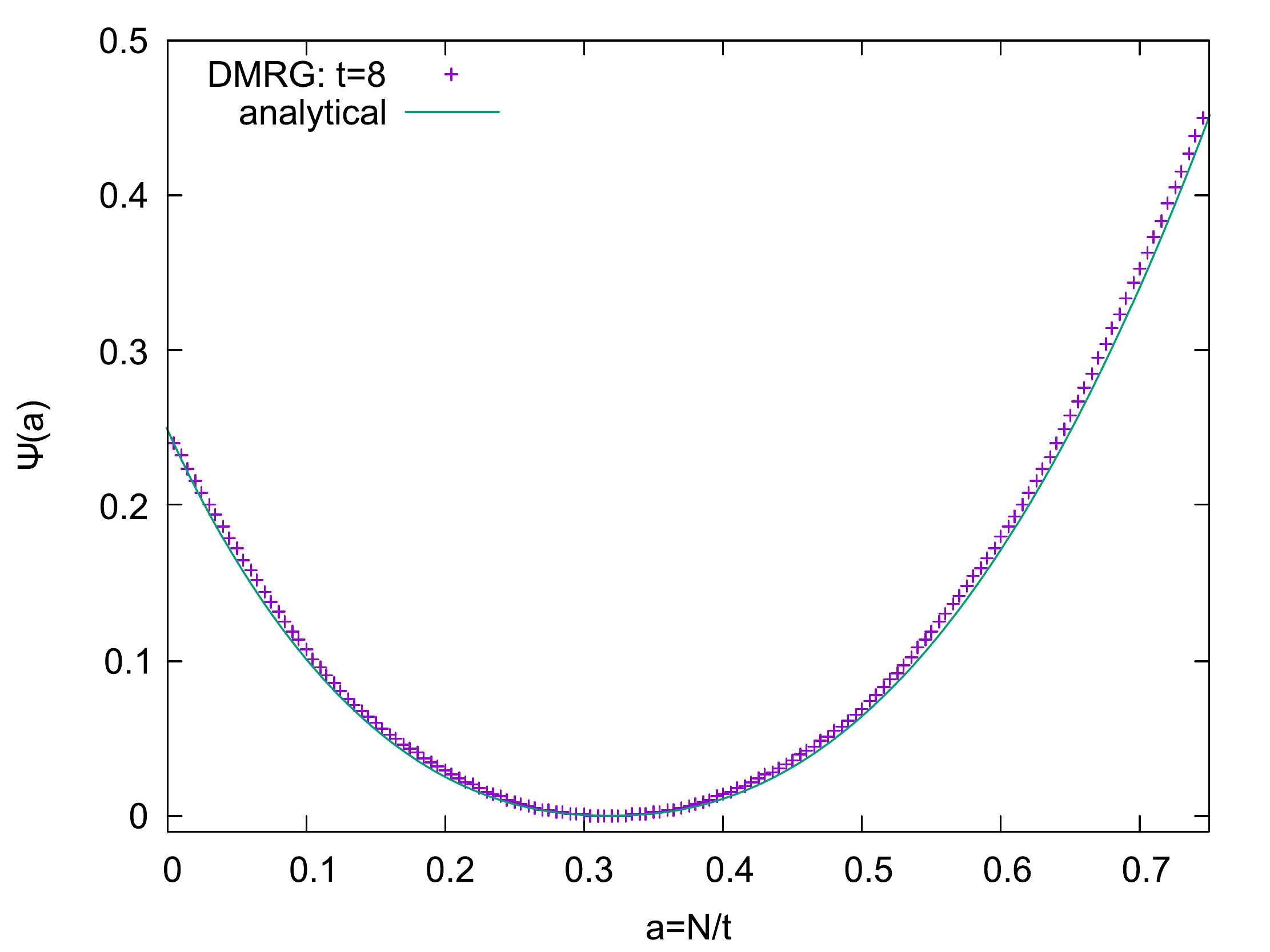}
\caption{Large deviation function for the XX model evolving from the domain wall initial state. The solid line is the theoretical prediction (see eq. (\ref{eq:LDF}) and eq. (\ref{eq:ator})). The dotted line is from DMRG calculation.}
\label{fig:LDF}
\end{figure}

We numerically verified the above formula of the large deviation function $ \Psi(a) $ by DMRG calculations, using the ITensor C++ library \cite{Itensor}. The result is shown in Fig. \ref{fig:LDF} exhibiting a striking agreement between analytical formula and the DMRG calculation.

We employed the matrix product states (MPS) time evolution method with matrix product operator (MPO) form of time-evolution operator $ e^{-it{H}} $\cite{PhysRevB.91.165112_longrangetDMRG} not only to simulate the real dynamics of the spin chain but also to obtain the cumulant generating function. For general information about MPS and MPO in DMRG, see for instance the reviews \cite{SCHOLLWOCK201196DMRGMPS,schollwock2011density}.

The large deviation function $ \Psi(a) $ can be calculated from the cumulant generating function of the integrated spin current \cite{TOUCHETTE20091}. However, as we have seen, the tail of the distribution has an exponent denoted by time squared $t^2$. This decay rate prevents us from applying the usual formula in large deviation theory.
For adapting our situation to the form available for the large deviation formalism, let us modify the moment generating function $\chi(\lambda,t)$.
Noting that $\lambda$ is just a parameter, we change the parameter $\lambda$ to $\lambda t$ and write $\chi(\lambda t,t)$ as $\tilde{\chi}(\lambda,t)$ after evaluating $\braket{e^{\lambda N(t)}}$.
In this way, we can obtain the large deviation function from the moment generating function :
\begin{equation}\label{eq:DMRG-definitionOfCGF}
\tilde{\chi}(\lambda) := \lim_{t\rightarrow\infty}\frac{1}{t^2}\ln\left\langle e^{\lambda t N(t)} \right\rangle,
\end{equation}
via the Legendre-Fenchel transform:
\begin{equation}\label{eq:DMRG-rateFunctionFromCGF}
{\Psi}(a) := \sup_{\lambda\in \mathbb{R}}\left\{ \lambda a - \tilde{\chi}(\lambda) \right\}.
\end{equation}

Thus, our goal is to obtain the numerical data of the expectation value $ \left\langle e^{\lambda N(t)} \right\rangle $ at each $ t $ and $ \lambda $. This quantity can be recast into the Schr\"{o}dinger picture:
\begin{align}\label{eq:DMRG-schroedingerCGF}
\left\langle{e^{\lambda{N}(t)}}\right\rangle 
&= \braket{{\phi(t=0)}|{e^{it{H}}e^{\lambda{N}(0)}e^{-it{H}}}|{\phi(t=0)}}\notag\\
&= \braket{{\phi(t)}|{e^{\lambda{N}(0)}}|{\phi(t)}}\notag\\
&=  \braket{{\phi(t)}|{e^{\lambda{N}(0)/2}e^{\lambda{N}(0)/2}}|{\phi(t)}}.
\end{align}
The procedure to evaluate this value is the following:
\begin{enumerate}
	\item Calculate the state $  \ket{\phi(t)} $ as an MPS by cumulatively applying the time evolution MPO $ e^{-i\Delta t{H}} $, from an initial state $ \ket{\phi(t=0)} $ of interest. In this study, we chose the Domain Wall state in eq. (\ref{DW}) with finite length $2L$.
	\item Prepare the MPO form of the operator $ e^{\pm\Delta\lambda N(0)} $ as in the case of the time evolution operator $ e^{-i\Delta t{H}} $, and cumulatively apply it to the state $ \ket{\phi(t)} $ up to $ \pm\lambda/2 $, when the desired value is $ \pm\lambda $. It is required to generate the states $ e^{\pm\lambda{N}(0)/2}\ket{\phi(t)} $ separately from $ \ket{\phi(t)} $ in order to obtain the value of $ \chi(\lambda) $ within the interval $ [-\lambda, \lambda] $.
	\item Evaluate the norm of the resulting state $ e^{\pm\lambda{N}(0)/2}\ket{\phi(t)} $.
\end{enumerate}
We can access to only finite values of $ t $ and $ \lambda $ in this manner because it is based on successive application of time evolution operator with finite time step. The range of $ \lambda $ determine the upper and lower bound of \emph{the gradient of the large deviation function}. Additionally, accurate calculation of cumulant generating function for large $ t $ requires much more bond dimension---the number of the singular vectors preserved in MPS, whose corresponding singular values are greater than or equal to the cutoff $ \epsilon $---of the MPS $ \ket{\phi(t)} $ and much smaller time step to reach it than other calculations, such as spin expectation values and correlation functions. We evaluated the reliability of the result by comparing the return probability $ \left|\braket{\phi(0)| \phi(t)}\right|^2 $ with the analytical expression $ e^{-t^2/4} $ in eq. (\ref{eq:RP}), which corresponds to the value of the rate function at the origin: $ {\Psi}(0) $.

We set our calculation parameters as follows; the length of the system: $ 2L = 13+13 $, the final time and the time step: $ t = 8, \Delta t = 2\times 10^{-6} $, the upper limit of the parameter of the cumulant generating function and its step: $ \lambda=36, \Delta \lambda = 5\times 10^{-3} $, the cutoff of singular values for $ t $ and $ \lambda $: $ \epsilon_t = 1\times 10^{-28}, \epsilon_\lambda = 1\times 10^{-13} $.

For these parameter values, the range of $a$ which could be numerically evaluated was restricted to $0\le a<0.75$.
In the region, we find a good agreement between the theory and numerics in Fig. \ref{fig:LDF}.
In the same figure, we also observe some deviation as $a$ increases.
We expect that the agreement becomes better as $t$ becomes large.
For instance, in another numerical calculation for a larger time $t=10$, we observed a better agreement than $t=8$, but the range of available $a$ was narrower for the reason that we need more information of  $\lambda$ as $t$ becomes large.

\section{Conclusion}

In this work, we studied the fundamental XX model and investigated the large deviation property of the spin current starting from the domain wall initial condition for absolute zero temperature and perfect transmission.
For this setup, we found that the values of the rare event for spin current in the right and left tail have an exponential decay in the form $e^{-t^2\Psi(a)}$.
The time squared decay seems inconsistent with the expectation from the approximated Levitov-Lesovik formula which predicts the decay rate $t$, but this is expected to be a robust behavior in our settings.
In addition to the exponent, the analytical large deviation function is exactly obtained.
The large deviation function $\Psi(a)$ has a non-symmetric shape, which would be a reflection of many-body effects in a non-equilibrium situation.

From the point of view of the experiment, one can construct the XX spin chain \cite{duan2003controlling}, using bosonic atoms in an optical lattice and adjusting the parameter of the system through changing of the laser intensity or the Feshbach resonance.
By trapping the hard core bosons undergoing Bose-Einstein condensation only in a center of periodic lattice potential plus harmonic potential by optical laser beams \cite{bloch2008many}, the initial condition corresponding to our domain wall initial state could be prepared.
Indeed, characteristic density profile of the form $\braket{s_m(t)}+1/2=\pi^{-1}\arccos\xi$ with $\xi=m/t$ has been observed by L. Vidmar {\it et al.} as shown Fig.2 in \cite{vidmar2015dynamical}.
As discussed below the eq. (\ref{eq:defP}), our results apply to both ferromagnetic and antiferromagnetic coupling.
The antiferromagnet XX spin chain is realized by compounds such as $\mathrm{PrCl_3}$ \cite{goovaerts1984pseudospin} or $\mathrm{CsCoCl_3}$ \cite{mohan1983excitation}.
One can apply strong magnetic field to these materials for preparing the domain wall.
It would be interesting to measure the fluctuation of $N(t)$ in experiments to compare with our theoretical predictions. 

As a future prospect, analyzing the details of the dynamics of XX model with other initial conditions would be a most natural direction.
The analytical behavior of the large deviation for other models, in particular the XXZ spin chain with the same or other initial condition, are most interesting and challenging problems.
Furthermore, we would like to clarify the connection between this work and the description of GHD extended to higher fluctuation \cite{myers2018transport}.
It may also be interesting to see if our results and methods in this paper reveal some properties of the hard edge statistics, on the contrary.

\section{Acknowledgments}

The authors are grateful to T. Yoshimura for helpful discussions and comments, and to P. Forrester, G. M. Sch{\"u}tz, L. Vidmar for pointing out a few related reference.
T. S. is also grateful to M. Katori, T. Imamura, T. Nagao and T. Shirai for their comments.
H. M. also acknowledges the financial support from Advanced Research Center for Quantum Physics and Nanoscience, Tokyo Institute of Technology.
The work of T. S. is supported by JSPS KAKENHI Grant Numbers JP15K05203, JP16H06338, JP18H01141, JP18H03672, JP19K03665.
\\
\\
\noindent
{\it Note added.}---At the stage of the proofreading, we have realized that the large deviation function of the particle number in the hard edge is studied in a different fashion in \cite{holcomb2018random}. The expression of large deviation in \cite{holcomb2018random} seems different from ours but numerically looks almost the same.

\begin{appendices}
\section{Determinant Identity}
\label{sec:AppendixA}

In this appendix, we illustrate how we get the relation in eq. (\ref{eq:DBtoBes}).
The determinant expression of the moment generating function $\chi(\lambda,t)$ in eq. (\ref{eq:mgfwithC}) contains a single particle density  operator restricted in a specific area at time $t$ as a correlation kernel
\begin{equation}
C(t)=P\rho(t)P=PU(t,0)\rho(0)U^\dagger(t,0)P.
\end{equation}
Where $P$ is a projection operator onto the area that measurement would be performed, $U(t,0)$ is an unitary operator from 0 to $t$ generated by the single particle Hamiltonian in eq. (\ref{eq:singleH}) and $\rho(t)$ is a density operator of the system at time $t$. Expanding the cumulant generating function $\log{\chi(\lambda,t)}$ as
\begin{equation}
\ln\chi(\lambda,t)=-\sum_{m=1}^\infty\frac{(-1)^m}{m}\left(e^{i\lambda}-1\right)^m\mathrm{Tr}C^m
\label{eq:cumulantexp}
\end{equation}
enables us to handle the correlation kernel as a single particle operator. When we focus on the trace part in eq. (\ref{eq:cumulantexp}), that has the form
\begin{equation}
\mathrm{Tr} C^m=\mathrm{Tr}\left[PU(t,0)\rho(0)U^\dagger(t,0)P\cdots PU(t,0)\rho(0)U^\dagger(t,0)P\right].
\end{equation}

In the case we are considering, the initial condition is just given by projection operator onto the left chain
\begin{equation}
\rho(0)=\rho_\mathrm{DW}=P_L.
\end{equation}

Another projection operator $P$ for measurement of up spins which takes projection onto the site $\{\alpha+1,\alpha+2\dots\}$ are to be written as $P_\alpha$.

We now proceed to calculate the trace
\begin{equation}
\mathrm{Tr}C^m=\mathrm{Tr}\left[P_\alpha U(t,0)P_LU^\dagger(t,0)P_\alpha\cdots P_\alpha U(t,0)P_LU^\dagger(t,0)P_\alpha\right].
\label{TrC^m}
\end{equation}
The unitary operator $U(t,0)$ in the coordinate representation is in eq. (\ref{eq:timeevolution1}). To bring the propagator to that close to the form appearing in the Coordinate Bethe Ansatz \cite{schutz1997exact,Tracy2008} (c.f. Theorem 2.1 in \cite{Tracy2008} for $N=1$), let us move on to $z$ plane to carry out the sum over all the momentum $k$ of plane waves $e^{ikn}$. Then,
\begin{equation}
\begin{split}
U_{mn}(t,0) &= \int_{-\pi}^\pi\frac{dk}{2\pi}e^{ik(m-n)}e^{it\cos k}=i^{n-m}J_{n-m}(t) \\
 &= \oint_{|z|=1}\frac{dz}{2\pi iz}z^{m-n}e^{\frac{it}{2}(z+1/z)}=\oint_{|z|=1}\frac{dz}{2\pi iz}z^{n-m}e^{\frac{it}{2}(z+1/z)}.
\end{split}
\label{eq:timeevolution2}
\end{equation}
In the second equality, the integral representation of the Bessel function of the first kind \cite{olver2010nist}
\begin{equation}
J_n(t)=\frac{1}{2\pi}\int_0^{2\pi}e^{it\sin\theta-in\theta}d\theta
\end{equation}
is used.
The last equality is derived from the equivalence of the hopping rate to the nearest neighbor.
Instead of performing the Wick rotation, we replace $iz$ by $z$, which does not change the contour of $z$
\begin{equation}
U_{mn}(t,0)=\oint_{|z|=1}\frac{dz}{2\pi iz}(-iz)^{n-m}e^{\frac{t}{2}(z-1/z)}.
\label{eq:ForwardU}
\end{equation}
The general property of evolution operator,
\begin{equation}
U_{mn}^\dagger(t,0)=U_{nm}(0,t)=U_{nm}(-t,0),
\end{equation}
and changing the variable $z\rightarrow1/z$ lead to
\begin{equation}
U_{mn}^\dagger(t,0)=\oint_{|z|=1}\frac{dz}{2\pi iz}(iz)^{n-m}e^{\frac{t}{2}(z-1/z)}.
\label{eq:BackwardU}
\end{equation}
Substituting eq. (\ref{eq:ForwardU}) and eq. (\ref{eq:BackwardU}) into the trace expression in eq. (\ref{TrC^m}), we have
\begin{equation}
\begin{split}
\mathrm{Tr}C^m &= \sum_{n_1=\alpha+1}^\infty\sum_{n_2=-\infty}^0\cdots\sum_{n_{2m-1}=\alpha+1}^\infty\sum_{n_{2m}=-\infty}^0\oint_{|z_1|=1}\frac{dz_1}{2\pi iz_1}\cdots\oint_{|z_{2m}|=1}\frac{dz_{2m}}{2\pi iz_{2m}} \\
 &\quad \times \prod_{k=1}^{m}z_{2k-1}^{-n_{2k-1}+n_{2k}}z_{2k}^{n_{2k}-n_{2k+1}}\prod_{k=1}^{2m}e^{\frac{t}{2}(z_k-1/z_k)}.
\end{split}
\end{equation}
Note that $z_{2m+1}=z_1$, $n_{2k+1}=n_1$ and the like are assumed in this notation because of the periodicity originated from the trace.
By taking the summation over their indices $n_k~(k=0,\dots,2m)$, one can carry out the trace as
\begin{equation}
\begin{split}
\mathrm{Tr}C^m &= \oint_{|z|>1}\prod_{k=1}^{m}\frac{dz_{2k-1}dz_{2k}}{(2\pi i)^2}\sum_{n_{2k+1}=\alpha+1}^\infty\frac{e^{\frac{t}{2}(z_{2k-1}-1/z_{2k-1}+z_{2k}-1/z_{2k})}}{z_{2k}^{n_{2k+1}}z_{2k+1}^{n_{2k+1}}(z_{2k-1}z_{2k}-1)} \\
 &= \oint_{|z|>1}\prod_{k=1}^{m}\frac{dz_{2k-1}dz_{2k}}{(2\pi i)^2}\frac{e^{\frac{t}{2}(z_{2k-1}-1/z_{2k-1}+z_{2k}-1/z_{2k})}}{(z_{2k}z_{2k+1})^\alpha(z_{2k}z_{2k+1}-1)(z_{2k-1}z_{2k}-1)} \\
  &= \oint_{|z|>1}\prod_{k=1}^{2m}\frac{dz_k}{2\pi iz_k^\alpha}\frac{e^{\frac{t}{2}(z_k-1/z_k)}}{(z_kz_{k+1}-1)}.
\end{split}
\end{equation}
We use the following identity that Derrida and Gerschenfeld employed in \cite{Derrida2009},
\begin{equation}
\frac{e^{\frac{t}{2}(z_k-1/z_{k+1})}}{z_k-1/z_{k+1}}=\frac{1}{z_k-1/z_{k+1}}+\frac{1}{2}\int_0^te^{\frac{t'}{2}(z_k-1/z_{k+1})}dt'.
\end{equation}

Going back to the trace and switching the corresponding part so that this form appears in the integrand, we obtain
\begin{equation}
\begin{split}
\mathrm{Tr}C^m &= \oint_{|z|>1}\prod_{k=1}^{2m}\frac{dz_k}{2\pi iz_k^{\alpha+1}}\frac{e^{\frac{t}{2}(z_k-1/z_{k+1})}}{(z_k-1/z_{k+1})} \\
 &= \oint_{|z|>1}\prod_{k=1}^{2m}\frac{dz_k}{2\pi iz_k^{\alpha+1}}\left[\frac{1}{z_k-1/z_{k+1}}+\frac{1}{2}\int_0^te^{\frac{t_k}{2}(z_k-1/z_{k+1})}dt_k\right].
\end{split}
\end{equation}
Here we can see that
\begin{equation}
\begin{split}
 &\quad \oint_{|z_k|>1}\frac{dz_k}{2\pi iz_k^{\alpha+1}}\frac{1}{z_k-1/z_{k+1}}\frac{1}{2}\int_0^tdt_{k} e^{\frac{t}{2}(z_{k-1}-1/z_k)} \\
 &= \frac{1}{2}\oint_{|z_k|>0}\frac{dz_k}{2\pi iz_k^{\alpha+2}}\int_0^tdt_{k}\sum_{n=0}^\infty\frac{1}{(z_kz_{k+1})^n} e^{\frac{t}{2}(z_{k-1}-1/z_k)}.
\end{split}
\end{equation}
Similarly, we get
\begin{equation}
\begin{split}
 &\quad \oint_{|z_k|>1}\frac{dz_k}{2\pi iz_k^{\alpha+1}}\frac{1}{z_{k-1}-1/z_k}\frac{1}{z_k-1/z_{k+1}} \\
  &= \oint_{|z_k|>0}\frac{dz_k}{2\pi iz_k^{\alpha+2}}\sum_{m=0}^\infty\sum_{n=0}^\infty\frac{1}{(z_{k-1}z_k)^m}\frac{1}{(z_kz_{k+1})^n}.
\end{split}
\end{equation}

Since the integrands do not have any poles with respect to $z_k$, because of $\alpha>0$, these terms readily equal zero by the residue theorem.
Therefore, the calculation becomes rather simple as the poles which behave singularly no longer exist
\begin{equation}
\mathrm{Tr}C^m=\frac{1}{2^{2m}}\oint_{|z|>0}\int_0^t\prod_{k=1}^{2m}\frac{dz_kdt_k}{2\pi iz_k^{\alpha+1}}e^{\frac{t_k}{2}(z_k-1/z_{k+1})}.
\end{equation}
Rearranging the order of the exponent,
\begin{equation}
\mathrm{Tr}C^m=\frac{1}{2^{2m}}\oint_{|z|>0}\int_0^t\prod_{k=1}^{2m}\frac{dz_kdt_k}{2\pi iz_k^{\alpha+1}}e^{\frac{t_k}{2}z_k-\frac{t_{k-1}}{2}\frac{1}{z_k}}
\end{equation}
holds.
Changing the variables $z_k$ as $\sqrt{\dfrac{t_{k-1}}{t_k}}z_k$ leads to
\begin{equation}
\mathrm{Tr}C^m=\frac{1}{2^{2m}}\oint_{|z|>0}\int_0^t\prod_{k=1}^{2m}\frac{dz_kdt_k}{2\pi iz_k^{\alpha+1}}e^{\frac{\sqrt{t_kt_{k+1}}}{2}(z_k-1/z_k)}.
\end{equation}

Now let us recall that the Bessel function of the first kind of order $\alpha$ has a complex contour representation.
\begin{equation}
J_\alpha(t)=\oint_{|z|>0}\frac{dz}{2\pi i}\frac{e^{\frac{t}{2}(z-1/z)}}{z^{\alpha+1}}.
\end{equation}
Hence, we arrive at the following expression
\begin{equation}
\mathrm{Tr}C^m=\frac{1}{2^{2m}}\int_0^t\prod_{k=1}^{2m}dt_kJ_\alpha(\sqrt{t_kt_{k+1}}).
\label{eq:chainJ}
\end{equation}
Moreover one can integrated out $m$ variables alternatively, by taking new variables and using a formula related to two product of the Bessel functions \cite{olver2010nist}
\begin{equation}
\int zJ_{\mu}\left(az\right)J_{\mu}(bz)dz=\frac{z\left(aJ_{\mu+1}\left(az\right)J_{\mu}(bz)-bJ_{\mu}\left(az\right)J_{\mu+1}(bz)\right)}{a^{2}-b^{2}}.
\end{equation}
From this formula, we can proceed further
\begin{equation}
\begin{split}
 &\quad \frac{1}{4}\int_0^tJ_\alpha(\sqrt{t_{k-1}t_k})J_\alpha(\sqrt{t_kt_{k+1}})dt_k \\
 &= \frac{1}{2}\int_0^{\sqrt{t}}t_kJ_\alpha(\sqrt{t_{k-1}}t_k)J_\alpha(\sqrt{t_{k+1}}t_k)dt_k \\
 &= \frac{J_{\alpha+1}\left(\sqrt{t_{k-1}t}\right)\sqrt{t_{k-1}t}J_\alpha(\sqrt{t_{k+1}t})-J_\alpha\left(\sqrt{t_{k-1}t}\right)\sqrt{t_{k+1}t}J_{\alpha+1}(\sqrt{t_{k+1}t})}{2(t_{k-1}-t_{k+1})} \\
 &= \frac{J_\alpha\left(\sqrt{t_{k-1}t}\right)\sqrt{t_{k+1}t}J'_\alpha(\sqrt{t_{k+1}t})-J'_\alpha\left(\sqrt{t_{k-1}t}\right)\sqrt{t_{k-1}t}J_\alpha(\sqrt{t_{k+1}t})}{2(t_{k-1}-t_{k+1})} \\
 &= tK_\mathrm{Bes}^{(\alpha)}(t_{k-1}t,t_{k+1}t).
\end{split}
\end{equation}
We also used recurrence relation of $J_\alpha(t)$ in the third equality.
In this way, we obtain
\begin{equation}
\mathrm{Tr}C^m=\int_0^{t^2}\prod_{k=1}^{m}dt_kK_\mathrm{Bes}^{(\alpha)}(t_k,t_{k+1})=\mathrm{Tr}K_\mathrm{Bes}^{(\alpha) m}.
\end{equation}
In other words,
\begin{equation}
\mathrm{Tr}C^m=\int_0^t\prod_{k=1}^{m}dt_k\tilde{K}_\mathrm{Bes}^{(\alpha)}(t_k,t_{k+1})=\mathrm{Tr}\tilde{K}_\mathrm{Bes}^{(\alpha) m}.
\end{equation}
The proof has been completed.

\section{Derivation of the large deviation function}
\label{sec:AppendixB}

In this section we derive the formula for the rate function $\Psi(a)$ in eq. (\ref{eq:LDF}) and eq. (\ref{eq:ator}). The starting point is the eq. (\ref{eq:LDFforN}). At first, we decompose the rate function $\Psi(a)$ into four parts and name the first three terms as $I_1,I_2,I_3$,
\begin{equation}
\psi_I(\kappa)=\frac{1}{2}\int_0^\infty\frac{x}{2}\rho^\ast(x)dx-\frac{\mu}{2}\kappa-\frac{\eta}{2}-\frac{3}{4}=I_1+I_2+I_3-\frac{3}{4}.
\label{LDFforN}
\end{equation}
Each term is given by
\begin{align}
I_1 &= \frac{1}{2}\int_0^\infty\frac{x}{2}\rho^\ast(x)dx= \frac{1}{4\pi}\int_0^\infty\frac{x}{2}\sqrt{\frac{(\lambda_+-x)(\lambda_--x)}{x(t^2/4N^2-x)}}dx, \\
I_2 &= -\frac{\mu}{2}\kappa= \frac{1}{2}\kappa\int_{\min\{\lambda_-,t^2/4N^2\}}^{\max\{\lambda_-,t^2/4N^2\}} \left[G(x)-\frac{1}{2}\right]dx, \\
I_3 &= -\frac{\eta}{2}=-\frac{1}{2}\left\{\log{\lambda_+}-\frac{\lambda_+}{2}-\int_{\lambda_+}^\infty\left[G(x)-\frac{1}{x}\right]dx\right\}.
\label{eq:I_3}
\end{align}
The resolvent $G(x)$ in eq. (\ref{eq:G_pm(z)}) has a different sign depending on the value of the argument $x$.
From the definition of the resolvent $G(x)$ in eq. (\ref{eq:defG(z)}), when $r<1$,
\begin{equation}
G(x)=
\begin{cases}
\quad G_-(x), &\mathrm{for}\quad x\in (-\infty,0]\cup[\lambda_-,t^2/4N^2]\cup[\lambda_+,\infty), \\
\quad0, &\mathrm{for}\quad x \in (0,\lambda_-)\cup(t^2/4N^2,\lambda_+).
\end{cases}
\label{eq:Gr<1}
\end{equation}
Similarly, when $r>1$,
\begin{equation}
G(x)=
\begin{cases}
\quad G_+(x) &\mathrm{for}\quad x\in[t^2/4N^2,\lambda_-], \\
\quad G_-(x) &\mathrm{for}\quad x\in (-\infty,0]\cup[\lambda_+,\infty), \\
\quad0 &\mathrm{for}\quad x \in (0,t^2/4N^2)\cup(\lambda_-,\lambda_+),
\end{cases}
\label{eq:Gr>1}
\end{equation}
follows.
According to the formula in \cite{majumdar2012number} or from the direct calculation following \cite{friedman1954handbook}, we know that
\begin{equation}
I_1=\frac{1}{4}\left[1+\frac{t^2}{4N^2}(1-r)\left(1-\frac{t^2}{4N^2}\right)\right].
\label{I_1}
\end{equation}
The expression above has nothing to do with the order of $r,1$ on $[0,\infty)$.

For the other two terms $I_2,I_3$, we expand those into the series of $1/N$.
In the case of $r<1$,
\begin{equation}
\begin{split}
I_2 &= -\frac{n}{4N}\int_{\lambda_-}^{t^2/4N^2}dx \sqrt{\frac{x-\lambda_-}{x(t^2/4N^2-x)}}\sqrt{\lambda_+}\left(1-\frac{x}{2\lambda_+}+\cdots\right) \\
 &= \frac{n}{4N}2\sqrt{\lambda_-\lambda_+}\left[K\left(1-\frac{t^2/4N^2}{\lambda_-}\right)-E\left(1-\frac{t^2/4N^2}{\lambda_-}\right)\right]+\mathcal{O}\left(\frac{1}{N^3}\right) \\
 &= \frac{n}{2N^2}t\sqrt{r}\left[K\left(1-\frac{1}{r}\right)-E\left(1-\frac{1}{r}\right)\right]+\mathcal{O}\left(\frac{1}{N^3}\right),
\end{split}
\end{equation}
where $K$ and $E$ are the elliptic integrals defined in eq. (\ref{eq:KandE}).
In the first equality, we used, if $a<b$,
\begin{equation}
\begin{split}
\int_a^b\sqrt{\frac{x-a}{x(b-x)}}dx &= \int_0^{b-a}\frac{x}{\sqrt{x(x+a)(b-a-x)}}dx \\
 &=  2\int_0^{\sqrt{b-a}}\frac{x^2}{\sqrt{(x^2+a)(b-a-x^2)}}dx \\
 &= 2\int_0^1\frac{(b-a)x^2}{\sqrt{(1-x^2)\left[a+(b-a)x^2\right]}}dx \\
 &= -2\sqrt{a}\int_0^1\left[\frac{1}{\sqrt{1-x^2}\sqrt{1-(1-b/a)x^2}}-\frac{\sqrt{1-(1-b/a)x^2}}{\sqrt{1-x^2}}\right]dx \\
 &= -2\sqrt{a}\left[K\left(1-\frac{b}{a}\right)-E\left(1-\frac{b}{a}\right)\right],
\end{split}
\end{equation}
and the definition of $r$ in eq. (\ref{defofr}).
In the case of $r>1$, following to the previous case, we obtain a similar expression,
\begin{equation}
\begin{split}
I_2 &= \frac{n}{4N}\int_{\lambda_-}^{t^2/4N^2}dx \sqrt{\frac{\lambda_--x}{x(x-t^2/4N^2)}}\sqrt{\lambda_+}\left(1-\frac{x}{2\lambda_+}+\cdots\right) \\
 &= \frac{n}{4N}2\sqrt{\lambda_-\lambda_+}\left[E\left(1-\frac{t^2/4N^2}{\lambda_-}\right)-K\left(1-\frac{t^2/4N^2}{\lambda_-}\right)\right]+\mathcal{O}\left(\frac{1}{N^3}\right) \\
 &= \frac{n}{2N^2}t\sqrt{r}\left[K\left(1-\frac{1}{r}\right)-E\left(1-\frac{1}{r}\right)\right]+\mathcal{O}\left(\frac{1}{N^3}\right).
\end{split}
\label{I_2}
\end{equation}

Next, we write the integral part in the third term $I_3$ in eq. (\ref{eq:I_3}).
\begin{equation}
\begin{split}
 &\quad \int_{\lambda_+}^\infty\left[G_-(x)-\frac{1}{x}\right]dx \\
 &= \int_{\lambda_+}^\infty\left[\frac{1}{2}-\frac{1}{2}\sqrt{\frac{(z-\lambda_+)(z-\lambda_-)}{z(z-t^2/4N^2)}}-\frac{1}{x}\right]dx \\
 &= \int_{\lambda_+}^\infty\left[\frac{1}{2}-\frac{1}{2}\left(1-\frac{\lambda_-}{2x}\right)\left(1+\frac{t^2}{4N^2}\frac{1}{2x}\right)\sqrt{1-\frac{\lambda_+}{z}}-\frac{1}{x}\right]dx+\mathcal{O}\left(\frac{1}{N^3}\right).
\end{split}
\label{eq:intinI3}
\end{equation}
When $x<1$, the square root of the form $\sqrt{1-x}$ can be expanded in terms of $x$ as Taylor series,
\begin{equation}
\sqrt{1-x}=1-\frac{x}{2}-\sum_{k=2}^\infty\frac{(2k-3)!!}{2^kk!}x^k.
\end{equation}
Substituting this formula in the above expression, the integrand in eq. (\ref{eq:intinI3}) is written as
\begin{equation}
\frac{1}{2}-\frac{1}{2}\left(1-\frac{\lambda_-}{2x}\right)\left(1+\frac{t^2}{4N^2}\frac{1}{2x}\right)\left[1-\frac{\lambda_+}{2x}-\sum_{k=2}^\infty\frac{(2k-3)!!}{2^kk!}\left(\frac{\lambda_+}{x}\right)^k\right]-\frac{1}{x}.
\end{equation}

Expanding this integrand and put together the contributions up to the order $\mathcal{O}(1/N^2)$, this becomes
\begin{equation}
\begin{split}
 & -\frac{1}{4x}\left(-\lambda_-+\frac{t^2}{4N^2}-\lambda_+\right)-\frac{1}{x}+\frac{1}{2}\sum_{k=2}^\infty\frac{(2k-3)!!}{2^kk!}\left(\frac{\lambda_+}{x}\right)^k \\
 &\quad +\frac{1}{4x}\left(\frac{t^2}{4N^2}-\lambda_-\right)\left[\frac{\lambda_+}{2x}+\sum_{k=2}^\infty\frac{(2k-3)!!}{2^kk!}\left(\frac{\lambda_+}{x}\right)^k\right]+\mathcal{O}\left(\frac{1}{N^3}\right).
\end{split}
\end{equation}
Performing the integration from $\lambda_+$ to $\infty$, we obtain
\begin{equation}
\begin{split}
 & \int_{\lambda_+}^\infty\left[\frac{1}{4}\sum_{k=1}^\infty\frac{(2k-1)!!}{2^k(k+1)!}\left(\frac{\lambda_+}{x}\right)^{k+1}+\frac{\lambda_+}{8x^2}\left(\frac{t^2}{4N^2}-\lambda_-\right)\right. \\
 &\quad\left.\frac{1}{8x}\left(\frac{t^2}{4N^2}-\lambda_-\right)\sum_{k=1}^\infty\frac{(2k-1)!!}{2^k(k+1)!}\left(\frac{\lambda_+}{x}\right)^{k+1} \right]dx+\mathcal{O}\left(\frac{1}{N^3}\right) \\
 &=\frac{\lambda_+}{4}\sum_{k=1}^\infty\frac{(2k)!}{4^kkk!(k+1)!}+\frac{1}{8}\left(\frac{t^2}{4N^2}-\lambda_-\right) \\
 &\quad +\frac{1}{8}\left(\frac{t^2}{4N^2}-\lambda_-\right)\sum_{k=1}^\infty\frac{(2k)!}{4^k[(k+1)!]^2}+\mathcal{O}\left(\frac{1}{N^3}\right).
\end{split}
\end{equation}
The two infinite series appearing in the first and the third term converge to the following values,
\begin{equation}
\sum_{k=1}^\infty\frac{(2k)!}{4^kkk!(k+1)!}=\log4-1,
\end{equation}
\begin{equation}
\sum_{k=1}^\infty\frac{(2k)!}{4^k[(k+1)!]^2}=3-2\log4.
\end{equation}

Now, we return back to the original form of $I_3$ and we can see that
\begin{equation}
\begin{split}
I_3 &= -\frac{1}{2}\left\{\log{\lambda_+}-\frac{\lambda_+}{2}-\int_{\lambda_+}^\infty\left[G_-(x)-\frac{1}{x}\right]dx\right\} \\
 &= -\frac{1}{2}\log{\lambda_+}+\frac{\lambda_+}{4}+\frac{\lambda_+}{8}(\log4-1)+\frac{1}{16}\left(\frac{t^2}{4N^2}-\lambda_-\right) \\
 &\quad +\frac{1}{16}\left(\frac{t^2}{4N^2}-\lambda_-\right)(3-2\log4)+\mathcal{O}\left(\frac{1}{N^3}\right) \\
 &= -\frac{1}{2}\log4-\frac{1}{2}\log\left(1+\frac{t^2/4N^2-\lambda_-}{4}\right)+\left(1+\frac{t^2/4N^2-\lambda_-}{4}\right) \\
 &\quad +\frac{1}{2}\left(1+\frac{t^2/4N^2-\lambda_-}{4}\right)(\log4-1)+\frac{1}{4}\frac{t^2/4N^2-\lambda_-}{4} \\
 &\quad +\frac{1}{4}\frac{t^2/4N^2-\lambda_-}{4}(3-2\log4)+\mathcal{O}\left(\frac{1}{N^3}\right) \\
 &= \frac{1}{2}+\frac{1}{4}\left[\frac{t^2}{4N^2}(1-r)\right]+\mathcal{O}\left(\frac{1}{N^3}\right).
\label{I_3}
\end{split}
\end{equation}
In the last equality, we used the definition of $\lambda_-$ in eq. (\ref{defofr}).
Combining all the terms in eq. (\ref{I_1}), (\ref{I_2}), and  (\ref{I_3}) evaluated separately into eq. (\ref{LDFforN}), multiplying $\psi_I$ by $2N^2$ and taking the large $N$ limit, we finally arrive at the expression for the large deviation function $\Psi(a)$ in eq. (\ref{eq:LDF}).
Here we can see that the term which has the order $\mathcal{O}(1)$ in $I_1,~I_3$ and the remaining constant $3/4$ in eq. (\ref{LDFforN}) are canceled.

In a similar way, evaluating the constraint in eq. (\ref{eq:numofevinI}) so that we do not leave the fictitious parameter $N$ in the large $N$ limit, we can obtain eq. (\ref{eq:ator}) as a form which does not include the parameter $N$.
Since the right hand side of eq. (\ref{eq:numofevinI}) has only $1/N$ dependence, we can ignore its higher correction smaller than $1/N$.

For $0\le r<1$, equivalently for $0\le a<1/\pi$, eq. (\ref{eq:numofevinI}) becomes
\begin{equation}
\begin{split}
\frac{at}{N} &= \frac{1}{2\pi}\int_0^{\lambda_-}\sqrt{\frac{(\lambda_+-x)(\lambda_--x)}{x(t^2/4N^2-x)}}dx \\
 &= \frac{1}{2\pi}\frac{t}{2N}\int_0^rdx\sqrt{\frac{r-x}{x(1-x)}}\sqrt{\lambda_+}\left(1-\frac{x}{2\lambda_+}\frac{t^2}{4N^2}-\cdots\right).
\end{split}
\end{equation}
Recalling that $\lambda_+$ has an expansion as $\lambda_+=4+\mathcal{O}(1/N^2)$ from eq. (\ref{eq:propofG}), and taking the large $N$ limit, we have,
\begin{equation}
\begin{split}
a &= \frac{1}{2\pi}\int_0^r\sqrt{\frac{r-x}{x(1-x)}}dx \\
 &= \frac{1}{2\pi}\int_0^1r\sqrt{\frac{1-x}{x(1-rx)}}dx \\
 &= \frac{1}{2\pi}\int_0^1r\sqrt{\frac{x}{(1-x)(1-r+rx)}}dx \\
 &= \frac{1}{\pi}\sqrt{1-r}\int_0^1\frac{rx/(r-1)}{\sqrt{(1-x)}\sqrt{1-rx/(r-1)}}dx \\
 &= \frac{1}{\pi}\sqrt{1-r}\left[E\left(\dfrac{r}{r-1}\right)-K\left(\dfrac{r}{r-1}\right)\right].
\end{split}
\label{eq:a(r)<t/pi}
\end{equation}

For $1<r$, equivalently for $1/\pi<a$, we can apply the same procedure and get the corresponding result as
\begin{equation}
a=\frac{1}{\pi}\sqrt{r}E\left(\dfrac{1}{r}\right).
\label{eq:a(r)>t/pi}
\end{equation}

\section{Constant in the Variance}
\label{sec:AppendixC}

In section \ref{sec:Variance}, we investigated the main contribution to the variance of the integrated spin current $\braket{N(t)^2}_c$ for large $t$.
In this appendix, we should determine the further correction up to $\mathcal{O}(1)$ order, by using the asymptotics of the Bessel kernel.

Making use of the expansion of the cumulant generating function $\log\chi(\lambda,t)$, we see
\begin{equation}
\begin{split}
\log\chi(\lambda,t) &= \sum_{n=1}^\infty\frac{(-1)^{n+1}}{n}(e^\lambda-1)^n\mathrm{Tr}\tilde{K}_\mathrm{Bes}^{(0)n} \\
 &= \sum_{m=1}^\infty\frac{\lambda^m}{m!}\sum_{n=1}^mW(m-1,n-1)\mathrm{Tr}\tilde{K}_\mathrm{Bes}^{(0)n}
\end{split}
\label{eq:CumulantExp}
\end{equation}
where $W(m,n)=(-1)^nn!S(m+1,n+1)$ is the (signed) Worpitzky number and $S(m,n)$ is the Stirling number of the second kind \cite{olver2010nist} (see chapter 26.8).
One of the exponential generating function of $S(m,n)$ \cite{olver2010nist} is given as follows,
\begin{equation}
\sum_{n=0}^\infty S(n,k)\frac{x^n}{n!}=\frac{(e^x-1)^k}{k!}.
\end{equation}
From the second coefficient, we have 
\begin{equation}
\braket{N(t)^2}_c=\mathrm{Tr}K_\mathrm{Bes}^{(0)}\left[1-K_\mathrm{Bes}^{(0)}\right]=\mathrm{Tr}\tilde{K}_\mathrm{Bes}^{(0)}\left[1-\tilde{K}_\mathrm{Bes}^{(0)}\right].
\end{equation}
This kind of trace is explicitly written down as follows by using eq. (\ref{eq:chainJ}), 
\begin{equation}
\begin{split}
\braket{N(t)^2}_c &= \frac{1}{2^2}\int_0^tdt_1\int_0^tdt_2J_0(\sqrt{t_1t_2})J_0(\sqrt{t_2t_1}) \\
 &\quad-\frac{1}{2^4}\int_0^tdt_1\cdots\int_0^tdt_4J_0(\sqrt{t_1t_2})J_0(\sqrt{t_2t_3})J_0(\sqrt{t_3t_4})J_0(\sqrt{t_4t_1}) \\
 &= \int_0^{\sqrt{t}}dt_1\int_0^{\sqrt{t}}dt_2t_1t_2J_0(t_1t_2)J_0(t_2t_1) \\
 &\quad-\int_0^{\sqrt{t}}dt_1\cdots\int_0^{\sqrt{t}}dt_4t_1t_2t_3t_4J_0(t_1t_2)J_0(t_2t_3)J_0(t_3t_4)J_0(t_4t_1).
\end{split}
\end{equation}
Moreover, using the representation of the Dirac delta function in terms of the Bessel function $J_{\nu}(t)$ \cite{olver2010nist},
\begin{equation}
\delta\left(x-y\right)=x\int_{0}^{\infty}tJ_{\nu}\left(xt\right)J_{\nu}\left(yt\right)\mathrm{d}t,
\end{equation}
we can unite the two multiple integral terms as
\begin{equation}
\begin{split}
\braket{N(t)^2}_c &= \int_0^{\sqrt{t}}dt_1\int_0^{\sqrt{t}}dt_2\int_0^{\sqrt{t}}dt_3t_1t_2J_0(t_1t_2)J_0(t_2t_3) \\
 &\quad\times\left[\delta(t_3-t_1)-\int_0^{\sqrt{t}}dt_4t_3t_4J_0(t_3t_4)J_0(t_4t_1)\right] \\
 &= \int_0^{\sqrt{t}}dt_1\int_0^{\sqrt{t}}dt_2\int_0^{\sqrt{t}}dt_3t_1t_2J_0(t_1t_2)J_0(t_2t_3)\int_{\sqrt{t}}^\infty dt_4t_3t_4J_0(t_3t_4)J_0(t_4t_1) \\
 &= \frac{1}{2^4}\int_0^tdt_1\int_0^tdt_2\int_0^tdt_3\int_t^\infty dt_4J_0(\sqrt{t_1t_2})J_0(\sqrt{t_2t_3})J_0(\sqrt{t_3t_4})J_0(\sqrt{t_4t_1}).
\end{split}
\end{equation}
if one integrates the variables $t_1$ and $t_3$, this leads to the double integral,
\begin{equation}
\begin{split}
\braket{N(t)^2}_c &= \int_0^tdt_2\int_t^\infty dt_4\tilde{K}_\mathrm{Bes}^{(0)}(t_2,t_4)\tilde{K}_\mathrm{Bes}^{(0)}(t_4,t_2) \\
 &= \int_0^tdx\int_t^\infty dy\frac{xy}{(x^2-y^2)^2}\left[J_0(x)yJ'_0(y)-J'_0(x)xJ_0(y)\right]^2.
\end{split}
\label{eq:exactVar}
\end{equation}
Applying asymptotic expansion of the Bessel function $J_{\nu}(t)$ and its derivative $J'_{\nu}(t)$ for a large argument \cite{olver2010nist},
\begin{equation}
\begin{split}
J_{\nu}(t)\simeq\sqrt{\frac{2}{\pi t}}\cos\left(t-\frac{1}{2}\nu\pi-\frac{1}{4}\pi\right), \\
J'_{\nu}(t)\simeq-\sqrt{\frac{2}{\pi t}}\sin\left(t-\frac{1}{2}\nu\pi-\frac{1}{4}\pi\right),
\end{split}
\end{equation}
we can perform the integration in eq. (\ref{eq:exactVar}) with elementary functions.
If one is interested in higher corrections, one could take subleading corrections of $J_\nu(t)$ and $J'_\nu(t)$ at this point.

The large $t$ behavior of this integral in eq. (\ref{eq:exactVar}) is reduced to the following integral,
\begin{equation}
\begin{split}
\braket{N(t)^2}_c &\simeq \frac{4}{\pi^2}\int_0^tdx\int_t^\infty dy\frac{1}{(x^2-y^2)^2} \\
 &\quad\times\left[x\sin\left(x-\frac{\pi}{4}\right)\cos\left(y-\frac{\pi}{4}\right)-y\sin\left(y-\frac{\pi}{4}\right)\cos\left(x-\frac{\pi}{4}\right)\right]^2 \\
 &= \frac{1}{\pi^2}\int_0^tdx\int_t^\infty dy\left[\frac{\cos(x+y)}{x+y}-\frac{\sin(x-y)}{x-y}\right]^2.
\end{split}
\end{equation}
Expanding the square in the integrand, we decompose the integral into three parts.
Let us name them as $J_1,J_2,J_3$ for simplicity,
\begin{equation}
\braket{N(t)^2}_c=J_1+J_2+J_3,
\end{equation}
where
\begin{align}
J_1 &:= \frac{1}{\pi^2}\int_0^tdx\int_t^\infty dy\frac{\cos^2(x+y)}{(x+y)^2}, \\
J_2 &:= -\frac{2}{\pi^2}\int_0^tdx\int_t^\infty dy\frac{\cos(x+y)\sin(x-y)}{x^2-y^2}, \\
J_3 &:= \frac{1}{\pi^2}\int_0^tdx\int_t^\infty dy\frac{\sin^2(x-y)}{(x-y)^2},
\end{align}
respectively.
We first integrate them with respect to $y$.
In the following, we are to use sine integral and cosine integral defined by \cite{olver2010nist}
\begin{equation}
\mathrm{Si}(z):=\int_0^z\frac{\sin x}{x}dx,
\end{equation}
\begin{equation}
\mathrm{Ci}(z):=-\int_z^\infty\frac{\cos x}{x}dx.
\end{equation}
For $J_1$, we have
\begin{equation}
\begin{split}
J_1 &= \frac{1}{\pi^2}\int_0^tdx\left[-\frac{\cos^2(x+y)}{x+y}-\mathrm{Si}(2x+2y)\right]_t^\infty \\
 &= \frac{1}{\pi^2}\int_0^tdx\left[\frac{\cos^2(x+t)}{x+t}+\mathrm{Si}(2x+2t)-\frac{\pi}{2}\right].
\end{split}
\end{equation}
Before the integration of $J_2$, let us rewrite the integrand to perform the integration easily as
\begin{equation}
\begin{split}
J_2 &= \frac{1}{\pi^2}\int_0^tdx\int_t^\infty\frac{1}{x}\left(\frac{1}{y+x}-\frac{1}{y-x}\right)\cos(y+x)\sin(y-x) \\
 &= \frac{1}{\pi^2}\int_0^t\frac{dx}{2x}\int_t^\infty dy\left(\frac{\sin2y-\sin2x}{y+x}-\frac{\sin2y-\sin2x}{y-x}\right) \\
 &= \frac{1}{\pi^2}\int_0^t\frac{dx}{2x}\int_t^\infty\frac{dy}{y+x}\left[\sin2(y+x)\cos2x-\cos2(y+x)\sin2x-\sin2x\right] \\
 &\quad -\frac{1}{\pi^2}\int_0^t\frac{dx}{2x}\int_t^\infty\frac{dy}{y-x}\left[\sin2(y-x)\cos2x+\cos2(y-x)\sin2x-\sin2x\right].
\end{split}
\end{equation}
This leads to
\begin{equation}
\begin{split}
J_2 &= \frac{1}{\pi^2}\int_0^tdx\frac{\cos2x}{2x}\left[\mathrm{Si}(2y+2x)-\mathrm{Si}(2y-2x)\right]_t^\infty \\
 &\quad -\frac{1}{\pi^2}\int_0^tdx\frac{\sin2x}{2x}\left[\mathrm{Ci}(2y+2x)+\mathrm{Ci}(2y-2x)+\log|y+x|-\log|y-x|\right]_t^\infty \\
 &= -\frac{1}{\pi^2}\int_0^t\frac{\cos2x}{2x}\left[\mathrm{Si}(2x+2t)+\mathrm{Si}(2x-2t)\right]dx \\
 &\quad +\frac{1}{\pi^2}\int_0^t\frac{\sin2x}{2x}\left[\mathrm{Ci}(2x+2t)+\mathrm{Ci}(2x-2t)+\log|t+x|-\log|t-x|\right]dx.
\end{split}
\end{equation}
The calculation of $J_3$ is similar to that of $J_1$.
Changing the variable as $y\rightarrow-y$ in the process of its calculation, we obtain
\begin{equation}
\begin{split}
J_3 &= \frac{1}{\pi^2}\int_0^tdx\left[\frac{\sin^2(x-y)}{x-y}-\mathrm{Si}(2x-2y)\right]_t^\infty \\
 &= \frac{1}{\pi^2}\int_0^tdx\left[-\frac{\sin^2(x-t)}{x-t}+\mathrm{Si}(2x-2t)+\frac{\pi}{2}\right].
\end{split}
\end{equation}
Since the second term $J_2$ vanishes for large $t$, we integrate the remaining two terms $J_1$, $J_3$,
\begin{equation}
\begin{split}
J_1+J_3 &= \frac{1}{\pi^2}\int_0^t\left[\frac{\cos^2(x+t)}{x+t}-\frac{\sin^2(x-t)}{x-t}+\mathrm{Si}(2x+2t)+\mathrm{Si}(2x-2t)\right]dx \\
 &= \frac{1}{\pi^2}\left[\frac{1}{2}\mathrm{Ci}(2x+2t)+\frac{1}{2}\mathrm{Ci}(2t-2x)+\frac{1}{2}\log|x+t|-\frac{1}{2}\log|t-x|\right. \\
 &\quad \left.+(x+t)\mathrm{Si}(2x+2t)+\frac{1}{2}\cos2(x+t)+(t-x)\mathrm{Si}(2t-2x)+\frac{1}{2}\cos2(t-x)\right]_0^t.
\end{split}
\end{equation}
Simplifying each term, we get
\begin{equation}
\begin{split}
J_1+J_3 &\simeq \frac{1}{2\pi^2}\left[\log t+\mathrm{Ci}(4t)-2\mathrm{Ci}(2t)+4t\mathrm{Si}(4t)-4t\mathrm{Si}(2t)\right. \\
 &\quad \left.+\cos4t-2\cos2t+\log4+\gamma+1\right],
 \end{split}
\end{equation}
where
\begin{equation}
\lim_{x\to0}\left[\mathrm{Ci}(x)-\log x\right]=\gamma=0.57721\cdots
\end{equation}
is the Euler-Mascheroni constant.
As the time $t$ becomes large, the following terms go to zero as
\begin{equation}
\lim_{t\to\infty}\left[4t\mathrm{Si}(4t)-4t\mathrm{Si}(2t)+\cos4t-2\cos2t\right]=0.
\end{equation}
An asymptotic property such that $\lim_{x\to\infty}\mathrm{Ci}(x)=0$ implies that both $\mathrm{Ci}(4t)$ and $2\mathrm{Ci}(2t)$ also vanish. 

In conclusion, we obtain the variance for large $t$ as
\begin{equation}
\braket{N(t)^2}_c\simeq\frac{1}{2\pi^2}\left(\log t+\log4+\gamma+1\right).
\label{eq:AppVar}
\end{equation}

\end{appendices}


\end{document}